\documentclass[twocolumn]{aastex63}
\pdfoutput=1 
\usepackage{amsmath,amstext}
\usepackage[T1]{fontenc}
\usepackage{apjfonts}
\usepackage{hyperref}
\usepackage[all]{hypcap}
\usepackage{url}
\usepackage{xspace}
\usepackage{color}

\usepackage{savesym}
\savesymbol{tablenum}
\usepackage[binary-units]{siunitx}
\restoresymbol{SIX}{tablenum}

\usepackage{booktabs}
\usepackage{tabularx}
\usepackage{multirow}
\usepackage{xcolor, colortbl}

\newcommand{\Spider}{\textsc{Spider}\xspace}
\newcommand{\spider}{\textsc{Spider}\xspace}
\newcommand{\Planck}{\textit{Planck}\xspace}
\newcommand{\planck}{\textit{Planck}\xspace}
\newcommand{\wmap}{\textit{WMAP}\xspace}
\newcommand{\WMAP}{\textit{WMAP}\xspace}
\newcommand{\muK}{\si{\micro\kelvin}\xspace}

\DeclareSIUnit[number-unit-product={}]\percent{\%}

\shorttitle{\Spider $B$-mode Results}
\shortauthors{\Spider Collaboration}

\begin{document}

\title{A Constraint on Primordial $B$-Modes from the First Flight of the \Spider Balloon-Borne Telescope}

\newcommand\ANL{High Energy Physics Division, Argonne National Laboratory, Argonne, IL, USA 60439}
\newcommand\CWRU{Physics Department, Case Western Reserve University, 10900 Euclid Ave, Rockefeller Building, Cleveland, OH 44106, USA}
\newcommand\Cardiff{School of Physics and Astronomy, Cardiff University, The Parade, Cardiff, CF24 3AA, UK}
\newcommand\UBC{Department of Physics and Astronomy, University of British Columbia, 6224 Agricultural Road,
Vancouver, BC V6T 1Z1, Canada}
\newcommand\Princeton{Department of Physics, Princeton University, Jadwin Hall, Princeton, NJ 08544, USA}
\newcommand\Caltech{Division of Physics, Mathematics and Astronomy, California Institute of Technology, MS 367-17, 1200 E. California Blvd., Pasadena, CA 91125, USA}
\newcommand\JPL{Jet Propulsion Laboratory, Pasadena, CA 91109, USA}
\newcommand\CITA{Canadian Institute for Theoretical Astrophysics, University of Toronto, 60 St. George Street, Toronto, ON M5S 3H8, Canada}
\newcommand\ASU{School of Electrical, Computer, and Energy Engineering, Arizona State University, 650 E Tyler Mall, Tempe, AZ 85281, USA}
\newcommand\UKZN{School of Mathematics, Statistics and Computer Science, University of KwaZulu-Natal, Durban, South Africa}
\newcommand\NITP{National Institute for Theoretical Physics (NITheP), KwaZulu-Natal, South Africa}
\newcommand\Imperial{Blackett Laboratory, Imperial College London, SW7 2AZ, London, UK}
\newcommand\Stockholm{The Oskar Klein Centre for Cosmoparticle Physics, Department of Physics, Stockholm University, AlbaNova, SE-106 91 Stockholm, Sweden}
\newcommand\Oslo{Institute of Theoretical Astrophysics, University of Oslo, P.O. Box 1029 Blindern, NO-0315 Oslo, Norway}
\newcommand\TorontoDunlap{Dunlap Institute for Astronomy and Astrophysics, University of Toronto, 50 St George Street, Toronto, ON M5S 3H4 Canada}
\newcommand\Toronto{Department of Astronomy and Astrophysics, University of Toronto, 50 St George Street, Toronto, ON M5S 3H4 Canada}
\newcommand\UIUCP{Department of Physics, University of Illinois at Urbana-Champaign, 1110 W. Green Street, Urbana, IL 61801, USA}
\newcommand\UIUCA{Department of Astronomy, University of Illinois at Urbana-Champaign, 1002 W. Green Street, Urbana, IL 61801, USA}
\newcommand\NRAO{National Radio Astronomy Observatory, Charlottesville, NC 22903, USA}
\newcommand\Michigan{Department of Physics, University of Michigan, 450 Church Street, Ann  Arbor, MI 48109, USA}
\newcommand\TorontoP{Department of Physics, University of Toronto, 60 St George Street, Toronto, ON M5S 3H4 Canada}
\newcommand\Hopkins{Department of Physics and Astronomy, Johns Hopkins University, 3701 San Martin Drive, Baltimore, MD 21218 USA}
\newcommand\Goddard{NASA Goddard Space Flight Center, Code 665, Greenbelt, MD 20771, USA}
\newcommand\APC{APC, Univ. Paris Diderot, CNRS/IN2P3, CEA/Irfu, Obs de Paris, Sorbonne Paris Cit\'e, France}
\newcommand\PennState{Department of Astronomy and Astrophysics, Pennsylvania State University, 520 Davey Lab, University Park, PA 16802, USA}
\newcommand\NIST{National Institute of Standards and Technology, 325 Broadway Mailcode 817.03, Boulder, CO 80305, USA}
\newcommand\Stanford{Department of Physics, Stanford University, 382 Via Pueblo Mall, Stanford, CA 94305, USA}
\newcommand\SLAC{ SLAC National Accelerator Laboratory, 2575 Sand Hill Road, Menlo Park, CA 94025, USA}
\newcommand\PrincetonEngineering{Department of Mechanical and Aerospace Engineering, Princeton University, Engineering Quadrangle, Princeton, NJ 08544, USA}
\newcommand\Fermilab{Fermi National Accelerator Laboratory, P.O. Box 500, Batavia, IL 60510-5011, USA}
\newcommand\KICPChicago{Kavli Institute for Cosmological Physics, University of Chicago, 5640 S Ellis Avenue, Chicago, IL 60637 USA}
\newcommand\Orsay{Institut d'Astrophysique Spatiale, Orsay, France}
\newcommand\MPI{Max-Planck-Institute for Astronomy, Konigstuhl 17, 69117, Heidelberg, Germany}
\newcommand\LAIM{Laboratoire AIM, Paris-Saclay, CEA/IRFU/SAp - CNRS - Universit\'e Paris Diderot, 91191, Gif-sur-Yvette Cedex, France}
\newcommand\WUSTL{Department of Physics, Washington University in St. Louis, 1 Brookings Drive, St.  Louis, MO 63130, USA}
\newcommand\MCSS{McDonnell Center for the Space Sciences, Washington University in St. Louis, 1 Brookings Drive, St.  Louis, MO 63130, USA}
\newcommand\Austin{Department of Physics, University of Texas, 2515 Speedway, C1600, Austin, TX 78712, USA}
\newcommand\McGill{Department of Physics, McGill University, 3600 Rue University, Montreal, QC, H3A 2T8, Canada}
\newcommand\StewardObs{Steward Observatory, 933 North Cherry Avenue, Tucson, AZ, 85721, USA}

\author{\spider Collaboration}
\noaffiliation{}

\author{ P.~A.~R.~Ade }
\affiliation{\Cardiff}

\author{ M.~Amiri }
\affiliation{\UBC}

\author{ S.~J.~Benton }
\affiliation{\Princeton}

\author{ A.~S.~Bergman }
\affiliation{\Princeton}

\author{ R.~Bihary }
\affiliation{\CWRU}

\author{ J.~J.~Bock }
\affiliation{\Caltech}
\affiliation{\JPL}

\author{ J.~R.~Bond }
\affiliation{\CITA}

\author{ J.~A.~Bonetti }
\affiliation{\JPL}

\author{ S.~A.~Bryan }
\affiliation{\ASU}

\author{ H.~C.~Chiang }
\affiliation{\McGill}
\affiliation{\UKZN}

\author{ C.~R.~Contaldi }
\affiliation{\Imperial}

\author{ O.~Dor{\'e} }
\affiliation{\Caltech}
\affiliation{\JPL}

\author{ A.~J.~Duivenvoorden }
\affiliation{\Princeton}
\affiliation{\Stockholm}

\author{ H.~K.~Eriksen }
\affiliation{\Oslo}

\author{ M.~Farhang }
\affiliation{\CITA}
\affiliation{\Toronto}

\author{ J.~P.~Filippini }
\affiliation{\UIUCP}
\affiliation{\UIUCA}

\author{ A.~A.~Fraisse }
\affiliation{\Princeton}

\author{ K.~Freese }
\affiliation{\Austin}
\affiliation{\Stockholm}

\author{ M.~Galloway }
\affiliation{\Oslo}

\author{ A.~E.~Gambrel }
\affiliation{\KICPChicago}

\author{ N.~N.~Gandilo }
\affiliation{\StewardObs}

\author{ K.~Ganga }
\affiliation{\APC}

\author{ R.~Gualtieri}
\affiliation{\ANL}

\author{ J.~E.~Gudmundsson }
\affiliation{\Stockholm}

\author{ M.~Halpern }
\affiliation{\UBC}

\author{ J.~Hartley }
\affiliation{\TorontoP}

\author{ M.~Hasselfield }
\affiliation{\PennState}

\author{ G.~Hilton }
\affiliation{\NIST}

\author{ W.~Holmes }
\affiliation{\JPL}

\author{ V.~V.~Hristov }
\affiliation{\Caltech}

\author{ Z.~Huang }
\affiliation{\CITA}

\author{ K.~D.~Irwin }
\affiliation{\Stanford}
\affiliation{\SLAC}

\author{ W.~C.~Jones }
\affiliation{\Princeton}

\author{ A.~Karakci }
\affiliation{\Oslo}

\author{ C.~L.~Kuo }
\affiliation{\Stanford}

\author{ Z.~D.~Kermish }
\affiliation{\Princeton}

\author{ J.~S.-Y.~Leung }
\affiliation{\Toronto}
\affiliation{\TorontoDunlap}

\author{ S.~Li }
\affiliation{\Princeton}
\affiliation{\PrincetonEngineering}

\author{D.~S.~Y.~Mak}
\affiliation{\Imperial}

\author{ P.~V.~Mason }
\affiliation{\Caltech}

\author{ K.~Megerian }
\affiliation{\JPL}

\author{ L.~Moncelsi }
\affiliation{\Caltech}

\author{ T.~A.~Morford }
\affiliation{\Caltech}

\author{ J.~M.~Nagy }
\affiliation{\WUSTL}
\affiliation{\MCSS}

\author{ C.~B.~Netterfield }
\affiliation{\Toronto}
\affiliation{\TorontoP}

\author{ M.~Nolta }
\affiliation{\CITA}

\author{ R. O\rq Brient}
\affiliation{\JPL}

\author{ B.~Osherson }
\affiliation{\UIUCP}

\author{ I.~L.~Padilla }
\affiliation{\Toronto}
\affiliation{\Hopkins}

\author{ B.~Racine }
\affiliation{\Oslo}

\author{ A.~S.~Rahlin }
\affiliation{\Fermilab}
\affiliation{\KICPChicago}

\author{ C.~Reintsema }
\affiliation{\NIST}

\author{ J.~E.~Ruhl }
\affiliation{\CWRU}

\author{ M.~C.~Runyan }
\affiliation{\Caltech}

\author{ T.~M.~Ruud }
\affiliation{\Oslo}

\author{ J.~A.~Shariff }
\affiliation{\CITA}

\author{ E.~C.~Shaw }
\affiliation{\UIUCP}

\author{ C.~Shiu }
\affiliation{\Princeton}

\author{ J.~D.~Soler }
\affiliation{\MPI}

\author{ X.~Song }
\affiliation{\Princeton}

\author{ A.~Trangsrud }
\affiliation{\Caltech}
\affiliation{\JPL}

\author{ C.~Tucker }
\affiliation{\Cardiff}

\author{ R.~S.~Tucker }
\affiliation{\Caltech}

\author{ A.~D.~Turner }
\affiliation{\JPL}

\author{ J.~F.~van~der~List }
\affiliation{\Princeton}

\author{ A.~C.~Weber }
\affiliation{\JPL}

\author{ I.~K.~Wehus }
\affiliation{\Oslo}

\author{ S.~Wen }
\affiliation{\CWRU}

\author{ D.~V.~Wiebe }
\affiliation{\UBC}

\author{ E.~Y.~Young }
\affiliation{\Stanford}
\affiliation{\SLAC}

\correspondingauthor{William C.~Jones}
\email{wcjones@princeton.edu}

\begin{abstract}
We present the first linear polarization measurements from the 2015
long-duration balloon flight of \Spider, an experiment designed to map the
polarization of the cosmic microwave background (CMB) on degree angular scales.
Results from these measurements include maps and angular power spectra from
observations of \SI{4.8}{\percent} of the sky at 95 and \SI{150}{\giga\hertz}, along with the results of
internal consistency tests on these data.  While the polarized CMB anisotropy
from primordial density perturbations is the dominant signal in this region of
sky, Galactic dust emission is also detected with high significance; Galactic
synchrotron emission is found to be negligible in the \Spider bands.  We employ
two independent foreground-removal techniques in order to explore the
sensitivity of the cosmological result to the assumptions made by each.  The
primary method uses a dust template derived from \Planck data to subtract the
Galactic dust signal.  A second approach, employing a joint analysis of \Spider
and \Planck data in the harmonic domain, assumes a modified-blackbody model for
the spectral energy distribution of the dust with no constraint on its spatial
morphology.  Using a likelihood that jointly samples the template amplitude and
$r$ parameter space, we derive \SI{95}{\percent} upper limits on the primordial
tensor-to-scalar ratio from Feldman--Cousins and Bayesian constructions, finding
$r<0.11$ and $r<0.19$, respectively.  Roughly half the uncertainty in $r$
derives from noise associated with the template subtraction.  New data at
\SI{280}{\giga\hertz} from \Spider's second flight will complement the \Planck polarization
maps, providing powerful measurements of the polarized Galactic dust emission.
\end{abstract}

\section{Introduction}
\label{sec:intro}

In the standard cosmological model ($\Lambda$CDM), the Universe consists of a blend of radiation, baryonic matter, cold dark matter, and a vacuum energy density
consistent with a
cosmological constant.
The observed structure in the Universe originates from primordial
fluctuations of matter and energy that grow through
gravitational instability.
 These perturbations evolve within a spacetime geometry that is spatially flat on the largest observed scales.
 This simple paradigm has proven to be in
remarkable agreement with the overwhelming majority of all
observational tests~\citep{peebles_annrev2012,planck2018_mission,planck2018_cosmology}.

Observational data place
stringent constraints on the properties of these primordial density fluctuations;
they must be predominantly adiabatic in nature, Gaussian-distributed, follow a nearly---but not quite---scale-invariant
spectrum, and encode correlations on scales larger than the
horizon during recombination.
Mechanisms to generate such fluctuations have been
proposed within the context of inflationary, bouncing, and cyclic models~\citep{guth82,starobinsky82,mukhanov82,hawking82,bardeen83,pdg2019,decadal_inflation2019,ijjas2018,ijjas2019,cook2020}.

In addition to the well-studied scalar perturbations, some early-Universe models---particularly inflationary models---predict a spectrum of
tensor perturbations, or primordial gravitational
 waves.
 Their amplitude is characterized by the dimensionless tensor-to-scalar ratio, $r$.\footnote{Throughout we specify $r$ at a scale of $k_0=0.05$\,Mpc$^{-1}$, and further assume a scale-invariant tensor spectrum ($n_t = 0$).  The six $\Lambda$CDM parameters are fixed to those of \cite{planck18_params}.}
The \Planck data combine precision measurements of the scalar fluctuations and the largest-scale CMB intensity fluctuations to constrain $r$ to be less than $r<0.10$~\citep{planck2018_inflation}.\footnote{This constraint relaxes to $r<0.16$ when excluding the low-$\ell$ data ($2\leq \ell \leq 29$) that include the temperature deficit.}

Local quadrupole anisotropies sourced by tensor fluctuations can
 also imprint a unique ``$B$-mode'' (curl) component to the polarization of the CMB at degree angular
 scales~\citep{kamionkowski_curl,zaldarriaga_bmode}.
 Though challenging to measure, this signature is relatively free of sample variance from the brighter scalar modes, and thus
 allows observational access to much smaller values of $r$.
The detection of
the signature of tensor fluctuations would bring remarkable new insights
into early-Universe physics. This scientific potential
has motivated an ambitious observational effort to search for the
signature of primordial gravitational waves in the polarization of
the CMB~\citep{Kamionkowski_review:2016,S4science}.

The \planck polarization data, spanning more than half of the full sky, constrain $r<0.158$ using limits on the $B$-mode contribution alone~\citep{tristram2021planck}.
Using $BB$ limits derived from observations of less than \SI{1}{\percent} of the full sky, the Keck team reports $r<0.072$~\citep{bk15}.
\planck measurements of the CMB intensity, the $E$-mode polarization, and lensing over more than half the full sky, together with the Keck $BB$ limits, improve the constraint to $r<0.056$~\citep{planck2018_inflation}.  In \cite{tristram2021planck} this same constraint is obtained using only \planck temperature and polarization data.  Combining the $B$-mode results from the Keck experiment with this re-analysis of the \planck polarization data, the same team reports a somewhat tighter constraint, $r<0.044$~\citep{tristram2021planck}.

As anticipated even prior to the \Planck results, any
cosmological $B$-mode signal is subdominant to the diffuse polarized emission
from our Galaxy along any line of sight~\citep{Fraisse}.
Current CMB observations must thus contend with modeling uncertainties associated with diffuse Galactic emission.
To date, the \Planck polarization data provide the most accurate estimate of polarized Galactic emission across the full sky \citep{Planck_2018_XI}.

In this paper we report results from the first flight of \spider, a balloon-borne instrument designed to measure the polarization of the CMB on degree angular scales.  
The paper is
organized as follows. After a brief description of the \spider
instrument in Section~\ref{sec:instrument} and observation strategy in Section~\ref{sec:flight},
we discuss the low-level
data processing leading up to maps of the sky in
Section~\ref{sec:processing}.  Section~\ref{sec:pse} presents two complementary angular
power spectrum estimators, while
Section~\ref{sec:nulls} discusses the
consistency tests performed with each of these estimators, and Section~\ref{sec:seb}
addresses sources of systematic error.
Results from several distinct methods of component separation are presented in Section~\ref{sec:fgs}, and Section~\ref{sec:likelihoods} provides constraints on cosmological parameters for each method.
The main conclusions and \spider's future prospects are summarized in Section~\ref{sec:conclusion}.

\section{The \spider instrument}
\label{sec:instrument}

The \spider payload consists of six monochromatic refracting telescopes housed within a single liquid helium cryostat, which is supported and pointed by a lightweight carbon fiber gondola.
Here we provide a brief overview of the payload design, and a more detailed description can be found in~\cite{RunyanSPIE2010,FilippiniSPIE2010,Rahlin2014,Gualtieri2018LTD}.

\subsection{Receivers}

\begin{table*}[t]
\centering
    \caption{Summary of instrumental parameters for the data used in this analysis.
    Band center and width are averages of per-detector measurements.
    Beam full-width at half-maximum is derived from a combined fit to all detectors in a given band.
    Noise-equivalent temperature is the quadrature average over all detectors used.
    Data used is the NET-weighted average of unflagged data in each channel, and
    is restricted to samples inside our sky mask (Section~\ref{sec:pse}) with
    hits-weighted $f_{sky}$ of \SI{3.9}{\percent}.
    Approximate map depths do not account for effects of filtering on signal-to-noise.  All sensitivities are reported in CMB temperature units.
    \label{table:det_params}}

    \begin{tabular}{cccccccc}
    \toprule
    \toprule
    ~ & Center & Width & FWHM & \# Det. & NET$_{tot}$ & Data Used & Map Depth\\[-4pt]
    Band & {[\si{\giga\hertz}]} & {[\si{\percent}]} & {[arcmin]} & Used & {[$\muK\sqrt{\si{\second}}$]} & {[days]} & {[\si{\micro\kelvin\cdot arcmin}]}\\
    \midrule
    \SI{95}{\giga\hertz} & 94.7 & 26.4 & 41.4 & 675 & 7.1 & 6.5 & 22.5 \\
    \SI{150}{\giga\hertz} & 151.0 & 25.7 & 28.8 & 815 & 6.0 & 5.6 & 20.4 \\
    \bottomrule
    \end{tabular}

\end{table*}

Each \spider receiver is an axisymmetric two-lens cryogenic refractor with a \SI{270}{\milli\meter} cold stop,
designed to minimize polarized systematics.
In each receiver, two high-density polyethylene lenses cooled to \SI{4}{\kelvin} focus light onto a \SI{300}{\milli\kelvin} focal plane.
The blackened cold stop and internal baffles surrounding the optics are cooled to \SI{1.6}{\kelvin} in order to reduce stray photon loading on the detectors.
A sapphire half-wave plate (HWP) mounted to a \SI{4}{\kelvin} flange skyward of each receiver's stop  is rotated to a new fixed orientation angle twice daily to provide polarization modulation~\citep{bryanSPIE,mechanism_paper}.
Each receiver views the sky through a series of reflective metal-mesh~\citep{metal_mesh_review} and lossy nylon filters to reduce infrared loading on the cryogenic system and detectors, as well as a thin ($\sim$\SI{3}{\milli\meter}) ultra-high-molecular-weight polyethylene (UHMWPE) vacuum window.
An appropriate single-layer anti-reflection coating, matched to the receiver\rq s band (95 or \SI{150}{\giga\hertz}), is attached to each side of the HWPs, lenses, vacuum windows, and relevant filters.

Each telescope focuses radiation onto four wafers (``tiles'') of antenna-coupled transition-edge sensors (TESs), fabricated at JPL~\citep{JPL_detectors}.
Each wafer is patterned with an array of polarimeter pixels, consisting of two inter-penetrating arrays of slot antennas (one for each perpendicular polarization mode).
This arrangement provides for an instantaneous measurement of total intensity and one of two linear polarization components.
A complete measurement of partial linear polarization---Stokes $I$, $Q$ and $U$ parameters---is obtained for each pixel through rotations of the HWP and the sky, which modulate the polarization angle~\citep{psb_methods}.
A microstrip feed network coherently couples optical power from these synthesized antennas through a band-defining lumped-element filter before dissipating the power incoherently on a thermally isolated island.
Each island supports two TESs with different critical temperatures, $T_c$, wired in series: a Ti sensor ($T_c\sim\SI{500}{\milli\kelvin}$) for science observations and an Al sensor ($T_c\sim\SI{1.3}{\kelvin}$) for laboratory testing.
The 512 (288) TESs of each \SI{150}{\giga\hertz} (\SI{95}{\giga\hertz}) focal plane are read out using a time-division SQUID multiplexing system~\citep{deKorte2003,Stiehl2011,Battistelli_MCE}.
The TESs and SQUIDs are housed within extensive magnetic shielding~\citep{RunyanSPIE2010}.

Table~\ref{table:det_params} summarizes the properties of all detectors used in the analysis presented in this paper.\footnote{In this paper, all temperatures used in reference to signal or noise are in units of $\Delta T_{CMB}$, the equivalent CMB fluctuation, in which the data are natively calibrated.}
This flight of \spider deployed a total of 2400 TESs. 
The channel counts in Table~\ref{table:det_params} account for intentionally dark (non-optical) TES channels, 
losses due to detector and readout performance, and the conservative channel cuts used in the present analysis. 
Notably, one of the three \SI{150}{\giga\hertz} receivers was excluded late in the analysis due to a null test failure (see Section~\ref{subsec:null_results}), but should be recoverable with future work.
Across the remaining five receivers, $\sim$\SI{80}{\percent} of TESs are used in this analysis.

\subsection{Cryogenics}
\spider's cryogenic system \citep{cryo_paper}, the largest yet deployed on a long-duration balloon flight, consists of two liquid
helium reservoirs: a 1284-L main tank and a 16-L superfluid tank.
The main tank is maintained at a pressure of roughly \SI{1}{\bar} during the flight, providing cooling power at $\sim$\SI{4}{\kelvin} for the receiver optics and the $^{3}$He sorption coolers.
The boil-off from the main tank flows through heat exchangers on each of two vapor-cooled shields, which intercept the radiative and conductive parasitic loads on the cryogenic system and cool the infrared filter stack.
The superfluid system provides cooling power at \SI{1.6}{\kelvin} to each telescope's $^{3}$He sorption cooler and internal optical baffles.
The superfluid tank fills continuously from the main tank through a capillary assembly, and is maintained at the ambient pressure of the altitude at float (about \SI{6}{\milli\bar}).
The superfluid system is pumped down on the ground, and maintained at low pressure during launch and ascent with a small diaphragm pump on the gondola.
The focal planes themselves are cooled to $\sim$\SI{300}{\milli\kelvin} by a dedicated $^{3}$He sorption cooler within each telescope.

\subsection{Gondola and Pointing System}
\label{sec:gondola_point}
The cryostat is supported within a lightweight carbon fiber gondola \citep{soler14}.
A reaction wheel and motorized pivot scan the gondola in azimuth, while a linear drive steps the cryostat in elevation \citep{shariff14}.
Absolute referencing of the payload orientation is provided by a suite of three star cameras: one attached to the cryostat and oriented along the boresight axis, the other two mounted to the outer gondola frame on a rotating table that allows them to track the sky during azimuthal scans.
Information from the star cameras is combined with that from GPS receivers, sun sensors, encoders, and gyroscopes to enable in-flight pointing and post-flight pointing reconstruction~\citep{gandilo14}.
Control and monitoring of the pointing and cryogenic systems is performed by a pair of redundant flight computers interfaced with the custom BLASTbus electronics \citep{benton14}.
A sun shield protects the instrument and optics during the 24-hour Antarctic summer daylight. Continuous electric power is provided by a \SI{2}{\kilo\watt} solar panel system, while various antenna arrays provide commanding, telemetry, and location information during the flight.

\section{Science Observations}
\label{sec:flight}
\spider was launched on January 1, 2015, from the NASA/NSF Long-Duration Balloon (LDB) facility near McMurdo Station, Antarctica.
All payload systems performed well throughout the flight, with the exception of a differential GPS unit failure that had no significant impact on flight operations or pointing reconstruction.
\spider's flight lasted 16.5~days at an average altitude of \SI{35}{\kilo\meter}.
The flight was terminated when cryogens were exhausted and the circumpolar wind system began to fail.
The payload touched down in a remote region of Ellsworth Land of West Antarctica.
Data drives and key flight hardware were recovered in February by personnel from the British Antarctic Survey;
a second team recovered the remainder of the instrument in November.
The payload optics and focal planes have subsequently been refurbished and upgraded in preparation for a second flight~\citep{shaw_spie}.

During an Antarctic LDB flight the Sun remains above the horizon at all times.  The accessible region of sky is therefore constrained by the need for the field center to remain roughly anti-solar and,
for CMB observations, to avoid the Galactic plane.
This favors a launch as early in the season as possible, since the anti-solar direction progresses to lower Galactic latitude over time.
\spider\rq s launch opportunity came relatively late in the Antarctic LDB season, pushing the field center toward the lower range of possible field centers.

\spider scanned in azimuth throughout the flight, with a sinusoidal speed profile peaking as  high as \SI{4}{\degree\per\second}.
The sinusoidal speed profile allowed smooth torque variations in the pivot and reaction wheel motors, without sustaining peak torque for long.
For the first two-thirds of the flight, \spider scanned a $\sim$\SI{75}{\degree} azimuthal range limited on either side by the Galaxy and the Sun.
In order to obtain more uniform coverage, the azimuthal range was reduced for the final third of the flight to cover the middle half of this range.
Scan turnarounds are separated by as long as \SI{36}{\second}, shortening to a little as \SI{22}{\second} for the narrower region later in the flight.
Small steps in elevation were made at every third scan turnaround, covering the full \SIrange{22}{50}{\degree} range upwards and downwards once per day.
A brief scan over the bright Galactic source RCW38 was used to confirm pointing in flight.

The in-flight pointing solution using only coarse sensors has an error of \SI{22}{\arcmin}~RMS, less than a \spider beam width and adequate for scan control.
The post-flight pointing reconstruction integrates gyroscopes between star camera solutions, and matches raw solutions of the boresight star camera to within \SI{0.9}{\arcmin}~RMS.
The relative pointing between the boresight camera and microwave detectors is calibrated with cross-correlation and deprojection methods described in Section~\ref{sec:deprojection}.

The half-wave plate is stepped in angle twice per sidereal day \citep{mechanism_paper, bryanSPIE}.
The nominal HWP angles are chosen to rotate each receiver between Stokes $Q$ and $U$ sensitivity every half day,
and to cover each rising and setting raster in $Q$ and $U$ with every detector on alternate days.
A total of eight discrete HWP angles, separated by \SI{22.5}{\degree} over a range of \SI{180}{\degree} are cycled through in an eight-day pattern to reduce
sensitivity to beam and HWP systematics \citep{bryan2010modeling, SpiderVpol}.  The HWP angles are measured with
a combination of absolute and relative encoders, providing an accuracy of \SI{0.1}{\degree}.

\spider's flight control system implements a number of autonomous watchdog routines that
monitor the quality of data returned from the detectors, perform limited corrective actions, and package compressed summary data
packets for return to the ground system.
Of particular note are the detector monitoring systems, described more fully in~\cite{Sasha_thesis}.
These use regular measurements of the TES differential resistance ($dV/dI$) from small \SI{2}{\hertz} square waves imposed on the TES bias lines; these are carried out for 2-second intervals
every fifth scan turnaround. These values are used to automatically identify channels that are superconducting or normal, have
accumulated a large DC offset, or have drifted significantly in TES resistance.
When the count of such anomalous channels grows large enough, the system
initiates a reset of the TES feedback loop or an adjustment of the TES bias.
Due to an unforeseen software race condition, this monitoring
system did not function for most receivers during the latter portion of the flight;
in practice this had little meaningful effect, given \spider's excellent detector performance stability.  Using both electrical and optical measurements of the temporal gain variations, the excursions on all timescales are found to be less than \SI{5}{\percent}, and not strongly correlated (see Section~\ref{sec:beams_cal}).

\section{Data Processing and Map Making}
\label{sec:processing}

Here we present an abbreviated discussion of \spider's low-level processing from raw data to calibrated maps and simulations of the sky.
More details can be found in \cite{Sasha_thesis, Anne_thesis, Ed_thesis}.

\spider's raw data consist of \SI{2.1}{\tera\byte} of time-ordered samples.
Bolometer and pointing data were recorded at \SI{119}{\hertz}, while a variety of
gondola and cryogenic performance parameters were recorded at reduced sample rates.
All data were recorded in-flight across multiple redundant drives.
Data from the six bolometer arrays and the flight system were synchronized
using data-valid clock signals and sequential counter values distributed from a single crystal clock system.

For a number of data processing operations,
samples are grouped into contiguous ``chunks'' approximately 10~minutes in length.
These evenly partition the periods between HWP angle steps and divide only at turn-arounds of the azimuthal scan.
Chunk length is a compromise between containing a sufficient number of samples for analysis tasks like estimating low-frequency noise, while remaining short enough that neighboring chunks have similar observing conditions and sky signal.
Ten minutes is long compared to the azimuthal scan period, but short relative to the timescale for changes in telescope elevation or cryogenic temperatures.
Similar chunk partitioning is also used to construct data subsets for power spectrum estimation (Section~\ref{sec:pse}).

\subsection{Timestream Flagging, Cleaning, and Filtering}

\label{sec:tod}

In addition to the expected Gaussian uncorrelated noise,
we observe two broad classes of correlated noise in
the \Spider timestream data: intermittent and quasi-stationary.
\emph{Intermittent} noise encompasses noise sources that appear to be
discretely on or off at any given time. The primary sources of
intermittent noise are the telemetry transmitters on board the
payload. The three Iridium transmitters in particular are active for
about two seconds at a time, operating asynchronously with periods
between 1 and 15 minutes.  Other sources of intermittent noise include
cosmic ray interactions in the detectors---discussed further
in \cite{spider_cray}---and various glitches or step discontinuities
due to the multiplexing readout. 
\emph{Quasi-stationary} noise
consists of non-astrophysical signals that are partially correlated
across the field of view and change very little over multiple azimuthal
scans. 
These have a peak-to-peak amplitude typically less than \SI{3}{\milli\kelvin_{CMB}}, and vary slowly over time.
It is believed that the majority of this contamination is sourced by
sidelobe pickup, primarily from the Earth's limb; its variation with elevation is not consistent with
signal from any residual atmospheric emission.  Additionally, RF-coupled interference was observed in some of the detector channels; one consequence of this is that a subset of detectors evidence a signal that is well-correlated with the
orientation of the reaction wheel, with couplings that vary in
strength both between warm readout electronics racks, and within them channel by channel.  During both pre-launch testing and the flight, one of the three readouts serving the \SI{150}{\giga\hertz} focal planes proved to be substantially more susceptible to these effects than the others.  As further discussed in Section~\ref{subsec:null_results}, the data from this focal plane contribute to null test failures, and are not included in any other part of this analysis.

Intermittent noise is mitigated by flagging of affected detector samples.
Such samples are tagged, 
replaced with constrained noise realizations, and excluded from map-making.
Step discontinuities arise intermittently in \spider's data, due primarily to transmitter interference and large cosmic ray interactions.
In addition to flagging the discontinuity itself, we adjust the data to eliminate the discontinuity using a linear fit to data before and after the event.
This procedure accounts empirically for cross-talk of the discontinuity among channels.
This ``stitching'' operation improves low-frequency noise significantly,
and simulations show that it has negligible effect on signal response.

Quasi-stationary noise is mitigated by time-domain filtering of this flagged data set, conducted at the full detector sample rate.
To reduce noise correlated with the reaction wheel, detector timestreams are binned according to the angle of the reaction wheel to form templates that are then subtracted.
The impact of this operation was checked using the full-flight time-domain simulations described in Section~\ref{sec:sims}, and found to have negligible effect on the astrophysical signal; this fit is thus performed only on data timestreams, but not on the large simulation ensembles.
To reduce low-frequency noise and pickup more broadly, the data are filtered between each scan turnaround by subtracting a fifth-order
polynomial fit to each detector's data as a function of azimuth.

\begin{figure*}
  \centering
  \includegraphics[width=\textwidth]{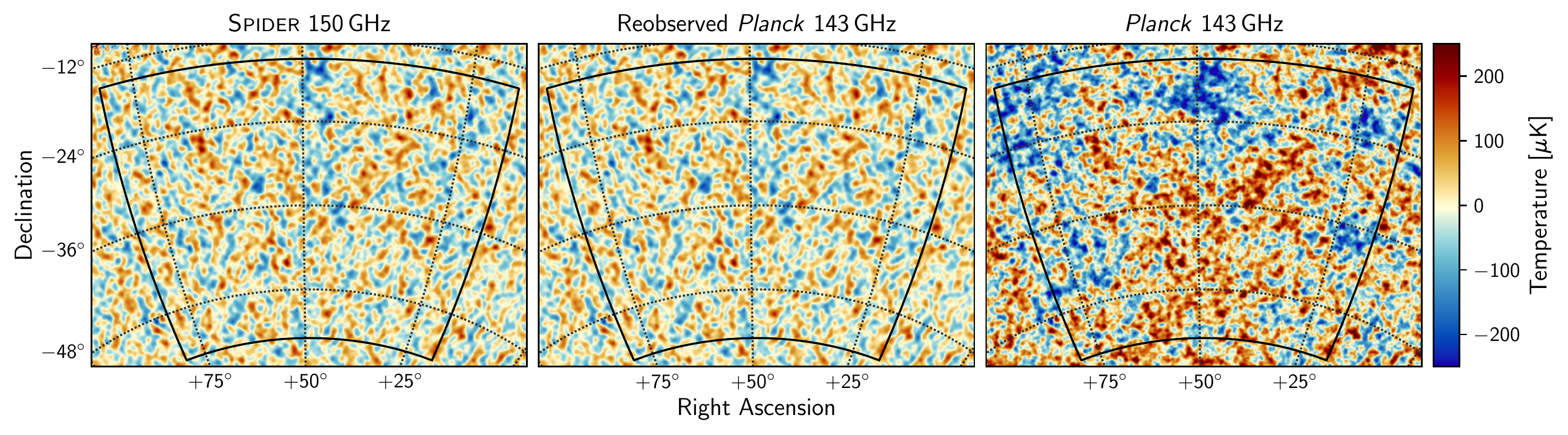}
  \caption{(\textit{left}) The total intensity map as observed by the \spider\ \SI{150}{\giga\hertz} receivers.  No additional filtering is applied to the maps beyond that in the timestream processing.  The black outline indicates the sky region used to compute power spectra, though the additional point source mask is not shown. (\textit{middle}) The \planck\ \SI{143}{\giga\hertz} map as re-observed using the \spider scan strategy and filtering, indicating strong agreement in the temperature signal.  (\textit{right})  The raw \planck\ \SI{143}{\giga\hertz} map, shown to illustrate the impact of \spider's scan strategy and filtering, which suppresses power at large angular scales.}
  \label{fig:tmap}
\end{figure*}

\begin{figure*}
  \centering
  \includegraphics[width=\textwidth]{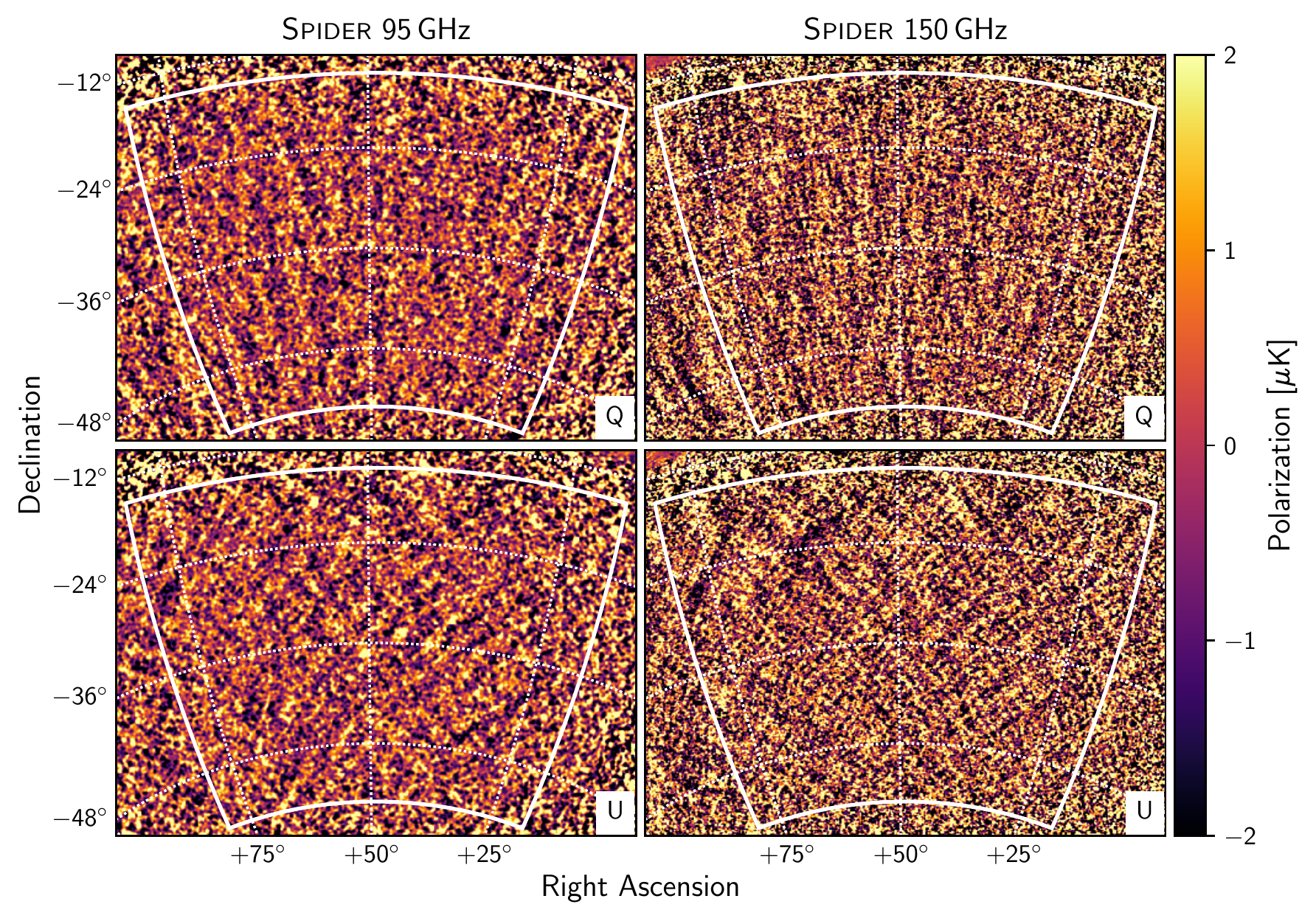}
  \caption{$Q$ and $U$ polarization maps as observed by \spider's 95 and \SI{150}{\giga\hertz} receivers.
    The maps have been smoothed with a \SI{10}{\arcminute} Gaussian for clarity.
    The temperature-to-polarization leakage from the map maker is subtracted (Section~\ref{sec:sims}), although this effect is not visible by eye.
    The dominant $E$-mode pattern of the cosmological signature is evident in the maps, although it is diluted by the Galactic signal.  The white outline indicates the sky region used to compute power spectra, though the additional point source mask is not shown.}
  \label{fig:pmap}
\end{figure*}

\begin{figure}
  \centering
  \includegraphics[width=\columnwidth]{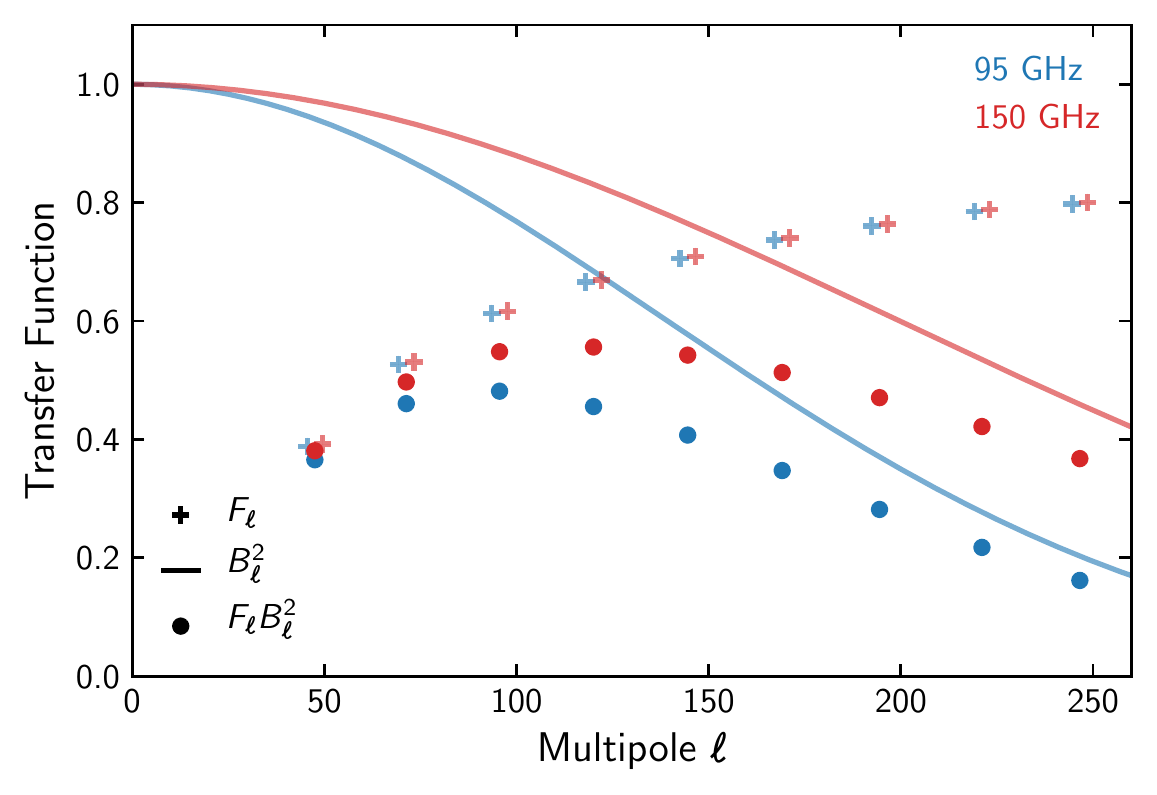}
    \caption{\spider's filter transfer function ($F_\ell$), beam window function ($B_\ell^2$), and total
    transfer function for 95 and \SI{150}{\giga\hertz}.  Quantities shown are the average of
    the $EE$ and $BB$ transfer functions, which are similar but are not assumed
    to be identical.  In this work, signal estimation is achieved using a simple
    binning of the data, which necessitates the filtering applied here.  This
    simple approach is driven by the desire to perform relatively efficient
    analysis of the signal and null tests in the time domain, but is unrelated
    to the atmosphere and does not represent a limitation of the stratospheric
    balloon platform.}
  \label{fig:fl_bl}
\end{figure}

The effects of scanning,
filtering, and flagging are determined by applying the entire analysis pipeline to an
ensemble of time-domain signal simulations.
The transfer functions due to filtering and beams, which derive from these simulations, are shown in Figure~\ref{fig:fl_bl} and are discussed further in Section~\ref{sec:pse}.
The primary effect of filtering is a suppression of power on large angular scales.
Because the typical scan speed and direction vary across the sky, the effect of
filtering is both anisotropic and inhomogeneous, with greater suppression on the edges of the field of view.
This latter effect is not visibly evident in the temperature or polarization maps (Figures~\ref{fig:tmap} and~\ref{fig:pmap}), and the net impact of the filtering has been shown to be adequately modeled with a simple multipole domain transfer function.

Detectors with consistently high noise after cleaning are completely cut from the analysis.
A small fraction ($\ll \SI{1}{\percent}$) of entire azimuth scans are also
flagged for having too much residual noise after cleaning.
In total, when weighting data by their estimated noise level, complete detector cuts remove \SI{5}{\percent} of the data, and \SI{28}{\percent} of samples on remaining detectors are flagged, including periods of cryogenic recycling. All detectors in the \SI{150}{\giga\hertz} receiver most strongly affected by RF-coupled interference are discarded, resulting in a further reduction of \SI{22}{\percent} at that frequency (\SI{10}{\percent} overall). 
Finally, \SI{27}{\percent} of unflagged samples lie outside the sky mask of the present analysis (Section~\ref{sec:pse}).

\subsection{Detector Characterization}
\label{sec:det_char}

Both pre-launch and in-flight data are used to characterize instrumental parameters needed to
construct accurate temperature and polarization maps.
The pre-launch data include spectroscopic and polarimetric measurements.
In-flight data are used for the absolute calibration, to monitor gain fluctuations, and to refine pre-launch estimates of beam response and pointing offsets.

\subsubsection{Polarization Angle}
\label{sec:trpns}

The individual detector polarization angles were measured prior to launch with a rotating polarized
thermal source in the near-field of each receiver as described in~\cite{Nagy_thesis}.
The uncertainty on each measured detector angle is approximately \SI{0.5}{\degree}, which is better than \spider's target of \SI{1}{\degree}~\citep{Fraisse}.
These measured angles are used directly by the map maker, with no correction applied based on the flight data.

\subsubsection{Frequency Response}
\label{sec:fts}

The frequency response of each detector was measured prior to launch with a custom high-throughput Fourier Transform Spectrometer (FTS) mounted on top of the cryostat.
Since the hot thermal source did not illuminate the full telescope solid angle, an actuated mirror steered the output over the full field of view.
\SI{95}{\percent} of all detectors used in the science analysis were measured with band center and band width accurate to \SI{1}{\giga\hertz}.
Measurements of the band centers and widths at different HWP angles and output mirror positions are consistent within errors.
Further details are provided in \cite{Anne_thesis}.

These per-detector measurements are not used directly in making maps, and this could result in leakage of spectrally mismatched temperature signals into polarization.
Instead, a null test is constructed splitting detectors with high and low band centers (Section~\ref{subsec:jackknives}).
Since no difference is detected, we conclude that the \spider data, including mitigation from HWP and sky rotation, have negligible leakage from bandpass mismatch.

\subsubsection{Calibration and Beam}
\label{sec:beams_cal}

\Spider's absolute calibration is derived by cross-calibrating degree-scale power with \Planck temperature anisotropy data at 100 and \SI{143}{\giga\hertz}\footnote{Throughout this paper we use release 3.01 of the \planck~HFI maps \citep{planck18_hfi}}.
This procedure finds the absolute calibration factor and parameterized beam model that minimizes the difference with the \Planck temperature spectra at a per-detector level in the range $100 < \ell < 275$ ($100 < \ell < 375$) for the 95\,(150)\,GHz frequency band. The absolute calibration is obtained by
finding the scalar, $c$, that minimizes
\begin{equation}
\sum _{\ell = \ell_1} ^{\ell_2} R_\ell \equiv \sum _{\ell = \ell_1} ^{\ell_2} \left| c\frac{\widehat{C}_{\ell}^{TT}}{\widehat{C}_{\ell, \mathrm{ref}}^{TT}} \frac{b_\ell^{\mathrm{Planck}}}{b_\ell^{\mathrm{SPIDER}}} - 1 \right|\,,
\end{equation}
where $\ell_1 = 100$ and $\ell_2 = 275 \:(375)$ for the 95\,(150)\,GHz frequency band. We use $\widehat{C}_{\ell,
  \mathrm{ref}}^{TT}$ to represent a temperature power spectrum
calculated using maps obtained from re-scanning the \Planck half-mission reference maps while $\widehat{C}_{\ell}^{TT}$ is calculated from single-detector
maps cross-correlated with a \Planck half-mission map.
The beam transfer functions, $b_\ell ^{\mathrm{Planck}}$
and $b_\ell ^{\mathrm{SPIDER}}$, quantify the relative sensitivity the
\Planck and \Spider spatial response as a function of multipole.

We use a simple Gaussian beam model, $b_\ell ^{\mathrm{SPIDER}}$, to extend that calibration to other angular scales included in our analysis; this extrapolation is small, and primarily to larger angular scales.
Each \spider telescope is fit with a single common beam model, which is then used to
determine an independent calibration factor for each individual detector.

Various consistency tests show that our analysis is not sensitive to a more physically motivated beam model, as significant deviations from a simple Gaussian are only evident below multipoles used in constructing bandpowers.
Using beam models informed by physical optics simulations, we have quantified the potential bias in our absolute calibration on the largest angular scales caused by the Gaussian beam model assumption.
At most, this results in a \SI{5}{\percent} bias in the beam transfer function in the lowest bin ($33 \leq \ell \leq 57$), which has a negligible impact on results.
The ``Inner/Outer Focal Plane Radius'' null test (Section~\ref{subsec:jackknives}) shows no detectable difference between the detectors expected to be the best and worst matches to our beam model.

We further explore the possibility of time-varying detector calibrations in several ways.
We use our regular TES resistance measurements (Section~\ref{sec:flight}) to generate a rough proxy for small changes in the TES bias state, and hence responsivity.
When TES monitoring is not available late in the flight, we use the average level of the TES current to calibrate a similar proxy.
We find that these estimates are consistent with one another.
Additionally, we find that gain excursions on all timescales are less than \SI{5}{\percent}, and are not strongly correlated.
While we correct our timestreams for an interpolated version of our TES-monitoring gain,
simulations and null tests show that this has negligible impact on our analysis (see Figure~\ref{fig:bb_residuals}).

Time domain simulations are used to quantify errors in the absolute calibration and effective beam width on both per-detector and full focal plane bases.
Statistical error in the determination of those parameters is caused by noise in the \Spider and \Planck data, both of which are incorporated in our simulations with appropriate noise models.
At the telescope level, statistical error is relatively small because of the data's high signal-to-noise ratio.
For example, the fractional error in the per-telescope beam transfer function, $b_\ell$, at degree angular scales, is approximately \SI{0.1}{\percent}.
The likelihood analysis described in Section~\ref{sec:likelihoods} incorporates a model of the statistical beam error.
Some potential systematic effects are also investigated and found to be subdominant at our sensitivity level (see Section~\ref{sec:seb}).

\subsubsection{Pointing Offset}
\label{sec:deprojection}

Each detector's pointing relative to the boresight star camera solution,
averaged over the full flight,
is initially characterized by maximizing the cross-correlation between
single-detector \spider temperature maps and \Planck maps.
To reduce error, a model for each detector tile---allowing free translation, rotation, and plate scale---is fit to the individual detector offsets.

These initial pointing offsets are refined using a time-domain ``deprojection'' technique
based upon \citet{bicep2_systematics}.
The deprojection method involves fitting for perturbations in leading-order
beam systematics---calibration, pointing offset, width, and ellipticity---using
time-domain templates generated from \Planck temperature maps and their derivatives.
Unlike the BICEP2 implementation, we fit for perturbations, not between paired detectors,
but between each detector's data and a simulation thereof, which is generated by re-observing \planck maps using the Gaussian beam model and initial pointing estimates from cross-correlation.
The per-detector pointing offsets measured this way are consistent with the cross-correlation
results but have greater precision.
In addition to measuring the average pointing offset of each detector over the full flight, the average offset of all detectors in each 10~minute chunk is used to measure and correct a slow thermal/mechanical drift over the course of the flight relative to the boresight position estimated by the pointing sensors.

Deprojection fits are also used to measure per-detector calibration, beam width, and ellipticity.
The estimated calibrations are in good agreement with those found in Section~\ref{sec:beams_cal} but are less precise and therefore not used.
The beam width and ellipticity parameters are used in Section~\ref{sec:seb} to simulate the effect of systematics not accounted for in the focal plane average beam model.

\subsection{Map Making}

In the next stage of the analysis, the processed data are binned into a two-dimensional map of the microwave sky by combining detector signal timestreams with reconstructed pointing and polarization angles.
As previously described, the detector signals input to the map maker are flagged, cleaned, and filtered (Section~\ref{sec:tod}),
before having calibrations applied (Section~\ref{sec:beams_cal}).
The input pointing timestreams are constructed by combining the boresight pointing and HWP angles (Section~\ref{sec:flight}) with per-detector polarization angles (Section~\ref{sec:trpns}) and pointing offsets (Section~\ref{sec:deprojection}).
Maps are made for each receiver and then combined by frequency band.
The resulting maps use \texttt{HEALPix} pixelization\footnote{\url{https://healpix.sourceforge.io }}
with $N_{side} = 512$ ($\sim$\SI{6.9}{\arcminute} resolution).

The cleaning and filtering process makes noise in the data largely uncorrelated
among channels and over time, with an approximately diagonal noise covariance between detector samples.
This simplification allows the maps to be constructed with simple weighted sums in each pixel \citep{psb_methods}, which in turn makes it computationally feasible to simulate large ensembles of time-domain simulations that include all relevant aspects of the experiment.
The weights are inverse noise variances of the cleaned and filtered data,
which are estimated independently for each detector and 10-minute chunk of data.

\subsection{Simulated Maps}
\label{sec:sims}

In addition to processing \spider data, the map maker can be run on simulated data
using the same flagging, filtering, beams, pointing, and polarization angles as \spider.
The simulated data can include signal from an input sky map, random noise generated from a power spectral density, and/or various injected glitches and systematics.

Noise simulations are generated separately for each detector, and are derived from the power spectral density of signal subtracted timestreams, averaged over all 10-minute chunks of data.
As such, the fiducial noise model assumes the detector noise is stationary and uncorrelated over the course of flight.
This model has been found to overestimate the true map noise in the data by $\sim$\SI{15}{\percent}, due primarily to the asymmetric impact of high outliers in the sample; it is thus empirically recalibrated by pipelines that use the noise simulations (Sections~\ref{subsection:XFaster} and~\ref{sec:smica}).

\Planck temperature and polarization maps are one source of known signal, matching the \spider bands at 95 and \SI{150}{\giga\hertz}
with the similar \Planck frequency maps at 100 and \SI{143}{\giga\hertz}.
A power spectrum can also be used as a source, where Gaussian random realizations
are made with the \texttt{synfast} utility from \texttt{HEALPix}.
In order to simulate the effects of various instrumental properties and systematics,
channel parameter values (pointing, calibration, etc.) may be applied differently when
simulating timestreams than when binning those simulated timestreams into a map.

Temperature-to-polarization leakage generated by the map-making pipeline is estimated by simulated observation of a \Planck temperature-only map.
The resulting polarization in this simulation, which is primarily caused by filtering of the temperature signal, constitutes a bias and is subtracted from data in the map domain.
In the harmonic domain, this is an approximately \SI{0.05}{\micro\kelvin\squared} correction in both $EE$ and $BB$ in units of $\ell(\ell+1)C_\ell/(2\pi)$.
The frequency mismatch between the \spider and \planck bands is neglected and thus introduces a small error in this bias subtraction.
We have verified, using \texttt{Commander} \citep{planck_foregrounds} temperature foreground estimates, that this approximation results in a relative error in the final polarization power spectra that is below \SI{1}{\percent} for each of our multipole bins.
Additionally, simulations show that $E$-to-$B$ leakage is negligible, measuring at most \SI{3}{\percent} of the $B$-mode error.
It is nonetheless corrected for in the NSI pipeline, described in Section~\ref{subsection:NSI}.

\subsection{Maps}
\label{sec:maps}

Figure~\ref{fig:tmap} shows the temperature map observed by \spider's \SI{150}{\giga\hertz} channels;
the \SI{95}{\giga\hertz} map (not shown) is visually very similar, since the dominant structure is fully resolved at both frequencies.
The corresponding polarization maps (Stokes $Q$ and $U$) are shown in Figure~\ref{fig:pmap}.
The rectangular outline shown over the maps encloses the region used for estimating angular power spectra, which covers \SI{4.8}{\percent} of the sky.

\section{Power Spectrum Estimation}
\label{sec:pse}

In order to estimate the underlying power spectrum of the sky from \spider's maps,
we must efficiently account for the effects of finite and uneven sky coverage, distortions from TOD processing
and the map maker, and the complex impact of instrumental noise.
The latter is particularly challenging to model or measure with the required precision and accuracy for space- and balloon-borne experiments~\citep{psb_methods}.
The relatively short duration of the observations provides limited data redundancy, which poses a challenge for fully empirical noise models~\citep[\emph{e.g.},][] {Ade:2014xna}.

We have developed two parallel power spectrum estimation pipelines for processing \spider maps:
XFaster, a maximum likelihood estimator; and the simpler Noise Simulation Independent (NSI) pipeline.
Each pipeline begins with a set of maps constructed from independent subsets of \spider's data, from which we construct a set of cross-spectra.
XFaster uses four data subsets, each combining every fourth 10-minute chunk of data (the same chunks used for low-level processing in Section~\ref{sec:processing}).
The NSI algorithm benefits from having a larger number of cross-spectra; this pipeline thus works with 14 data subsets composed from interleaved 3-minute chunks, with the shorter chunk length chosen to prevent gaps in sky coverage.

A common sky mask is used for all the results in this paper.
The mask covers 1964 square degrees with uniform weighting, consisting of the 1992 square degree rectangle shown in Figures~\ref{fig:tmap} and~\ref{fig:pmap} with point sources removed.
The point source mask excludes \SI{1}{\degree} diameter circular regions around objects from the Planck compact object catalog \citep{planck_compact_sources}, plus a \SI{2}{\degree} region around the bright radio galaxy NGC~1316.
The 50 brightest sources in \spider's observation region are masked this way, though not all lie within the chosen rectangle.
Among a handful of simple mask options, this mask was the largest subset of the data that was well-conditioned and passed null tests.
This mask was established prior to the calculation of the signal power spectra in order to avoid potential bias.
Null test and signal power spectra are computed with both pipelines, with results shown in Section~\ref{sec:nulls}.

Each pipeline ultimately produces a spectrum and covariance matrix for $33 \leq \ell \leq 257$, binned into nine ``science'' bandpowers with an $\ell$ width of 25.
One lower ($8 \leq \ell \leq 32$) and two higher bandpowers are also computed for each pipeline in order to accurately account for their leakage into the nine bins used for cosmological analysis.
The bin starting at $\ell=8$ was found to contain residual systematic signal, and the bins above $\ell=257$ contribute little to the cosmological and foreground constraints; thus, they are excluded from the science bins.
Throughout the following, unless explicitly stated otherwise, the lowest or first bin refers to the first science bin, \emph{i.e.} that starting at $\ell=33$.
The details that distinguish the two pipelines are provided below.

\subsection{XFaster Pipeline}
\label{subsection:XFaster}
XFaster is a maximum likelihood estimator, built as a hybrid
of Monte Carlo estimators, such as MASTER \citep{master} and PolSpice
\citep{polspice}, and iterative quadratic estimators
\citep{xfaster_planck,xfaster_tegmark}.
It was developed to allow the application of a maximum likelihood estimator to maps made with disjoint masks.
XFaster is based on an algorithm originally written for analysis of
the BOOMERanG data set \citep{Netterfield2002,xfaster_boomerang}, and has since been one of the
estimators used in the \planck analysis~\citep{xfaster_planck}.
A number of new features were implemented for use in the \spider pipeline,
notably the ability
to calculate null spectra.
The main features are summarized here, with further
details left to a dedicated paper~\citep{xfaster_forward}.

XFaster iteratively solves for bandpower deviations from a fiducial
full-sky signal model using an approximation for the likelihood of
cut-sky $a_{\ell m}$ modes.
This signal model is constructed using the MASTER formalism~\citep{master}, in which the mode-mixing from the mask is computed analytically (including an $E$-$B$ mixing component), the beams are pre-computed as described in Section~\ref{sec:beams_cal}, and the filter transfer functions are estimated from an ensemble of 1000 $\Lambda$CDM simulations run through the full map-making pipeline (Section~\ref{sec:sims}).
In addition to the signal power, XFaster also estimates the instrumental noise from the auto- and cross-spectra of the input maps.
An ensemble of 1000 time-domain noise simulations is input to the
pipeline to provide a fiducial noise model. The noise model is itself
iteratively recalibrated by including deviations from the fiducial
model as parameters in the likelihood maximization alongside the signal bandpowers.
The noise residual parameters are modeled as the same for $EE$ and $BB$.

In this pseudo-$a_{\ell m}$ space, the full likelihood of the observed
data modes $\widetilde{\pmb{d}}$, given the signal model $\widetilde{\pmb{S}}$ and noise, is
approximated, for a single map and for a single spectrum, as
\begin{equation}
  \label{xf_cl_likelihood}
  -2 \ln L(\widetilde{\pmb{d}}|\widetilde{\pmb{S}})=\sum_{\ell} g_{\ell}(2 \ell+1) \left[\frac{\widehat{C}_{\ell}}{\widetilde{S}_{\ell}+\widetilde{N}_{\ell}}+\ln \left(\widetilde{S}_{\ell}+\widetilde{N}_{\ell}\right)\right]\,,
\end{equation}
where $\widehat{C}_\ell$ is the data pseudo-spectrum,
$\widetilde{S}_\ell$ is the estimated signal, and $\widetilde{N}_\ell$ is the
noise bias estimated from the mean of noise-only simulation ensembles.
This approximation assumes that the signal and noise components are
uncorrelated.
The likelihood is diagonalized by assuming that, by binning power into bandpowers of sufficient width,
the effect of correlations between
multipoles on the estimate is greatly reduced.
In practice, the components of Equation~\ref{xf_cl_likelihood} are
matrices of all cross-spectra among all maps used for the analysis.
The structure of the generalized likelihood is band-diagonal to account for
correlations between spectral combinations and overlapping maps.

The vector $g_\ell$ is a recalibration of the effective mode count;
this corrects the likelihood, and the Fisher matrix obtained from it,
for the effects of masking, filtering, and diagonalizations.
The recalibration is computed using an ensemble of signal-only simulations.
For null tests, we add noise to the signal simulations when estimating $g_\ell$; without noise, the simulated null spectrum would be exactly null and not allow for calibration.

The iterative estimate of bandpower deviations on the signal and noise model automatically produces the Fisher information matrix, whose inverse is the bandpower covariance.
Extensive simulations have been performed to ensure the XFaster estimator is unbiased and that the resulting covariance matrix is accurate.
These are discussed further in \cite{xfaster_forward}.

\subsection{NSI Pipeline}
\label{subsection:NSI}

The Noise Simulation Independent (NSI) pipeline was developed to
estimate statistical bandpower errors directly from the data.
It uses PolSpice \citep{polspice} to compute the cross-spectra of 14 temporally
independent maps at each observing frequency, generated from interleaved 3-minute data chunks.
All possible cross-spectra are constructed from the map ensemble (neglecting the auto-spectra), providing 91 at each single frequency (95$\times$\SI{95}{\giga\hertz} or 150$\times$\SI{150}{\giga\hertz}) and 196 with one map at each frequency (95$\times$\SI{150}{\giga\hertz}), for a combined total of 378 cross-spectra.
The bandpowers are estimated from the noise-weighted mean of all
cross-spectra. The associated statistical uncertainties are estimated from the distributions of these cross-spectra by computing
the standard error on the mean with jackknife resampling.
By using many cross-spectra of uncorrelated maps, the sensitivity of the NSI pipeline approaches that
of an auto-spectrum analysis, but the resulting bandpower estimates are not biased
by any mischaracterization of the noise auto-spectrum.
This methodology is similar to Xspect \citep{Xspect} and Xpol \citep{Xpol},
 and similar approaches have been used by several experiments including SPT \citep{lueker_SPT} and CLASS \citep{padilla_class} 
as well as for \spider's circular polarization analysis \citep{SpiderVpol}.

The NSI pipeline uses a two-dimensional ``transfer matrix'' to correct the power spectra for mode mixing and power attenuation from filtering.  This approach considers the leakage from a given multipole bin to all others, both within the same spectrum and between spectral types ($TT$, $EE$, $BB$).
By design, this procedure also includes a correction for the instrumental beam as well as any residual leakage effects induced by the cut-sky mask that are not corrected by the spectral estimator.  Though found to be negligible for \spider, it also includes a correction for $E$-to-$B$ leakage.
The transfer matrix is constructed from a simulation ensemble in which each simulated map (Section~\ref{sec:sims}) has a source spectrum with only one non-zero multipole bin, which is set to the value of the appropriate fiducial $\Lambda$CDM spectrum.  When the maps are processed with the NSI pipeline, the ratio of the output and input spectra encodes the leakage from that bin to all others.
Further discussion of this method and its impact on the recovered spectra is provided in \cite{transfer_forward}.

\section{Consistency Tests}
\label{sec:nulls}
\spider's data processing is designed to flag or filter out the dominant sources of systematic contamination in the time-ordered data, such as intermittent pickup and quasi-stationary noise.
In order to ensure that low-level residuals do not remain at the level of our sensitivity, we conduct two types of tests: null tests (described in Section~\ref{subsec:jackknives}), and simulated injections of modelable systematics (described in Section~\ref{sec:seb}).
We also discuss the consistency between the two power spectrum estimation pipelines in Section~\ref{subsec:est_consistency}.

\subsection{Null Tests}
\label{subsec:jackknives}
Null tests check for systematic noise residuals in the differences between pairs of maps, 
constructed from various splits of \spider's data by time period or detector set.
The pairs of maps are chosen to share common signal but to have independent noise, 
and to maximize the residuals from possible systematic effects within the data.
If the power spectra of these differences are consistent with statistical noise, then we have evidence that systematic errors probed by the splits do not significantly contaminate the maps.
These tests are performed separately for both the NSI and XFaster pipelines.

\subsubsection{Null Split Definitions}
\begin{figure}
  \centering
  \includegraphics[width=0.4\columnwidth]{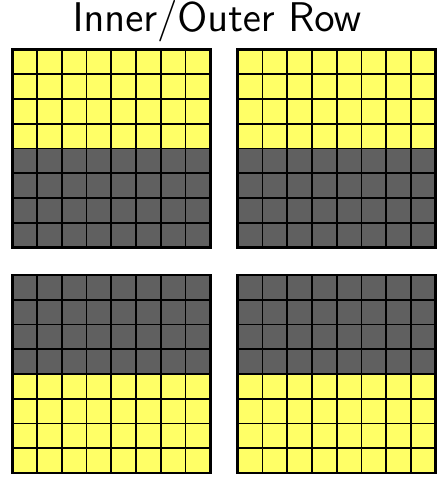}
  \includegraphics[width=0.4\columnwidth]{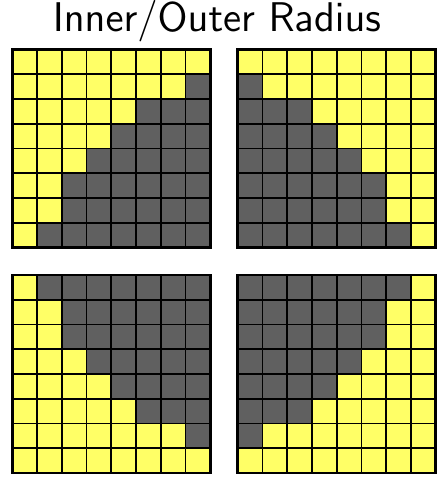}\\
  \vspace{0.05in}
  \includegraphics[width=0.4\columnwidth]{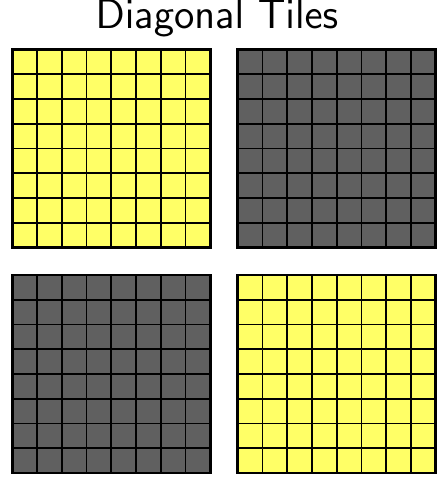}
  \includegraphics[width=0.4\columnwidth]{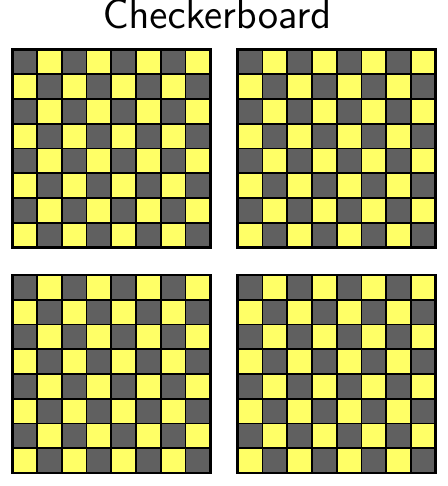}\\
  \vspace{0.05in}
  \includegraphics[width=0.4\columnwidth]{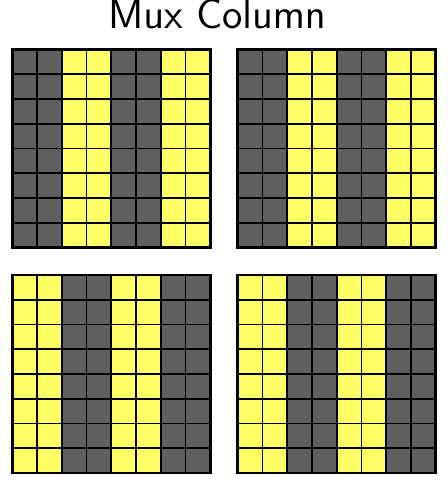}
  \caption{Physical representation of coordinate-based detector splits used for null tests.
     Each square represents a pixel, each of which contains two orthogonally polarized detectors.
     Focal plane maps shown are for \SI{150}{\giga\hertz}; \SI{95}{\giga\hertz} splits look similar but with 6$\times$6 grids of pixels per tile.}
  \label{fig:null_test_splits}
\end{figure}

The suite of null splits is listed below.
The first five splits are based on channel location within the focal plane, as illustrated in Figure~\ref{fig:null_test_splits}. 
Two further splits employ alternate divisions of the detectors: by pointing relative to payload azimuth (dependent on the orientation of each receiver about its boresight axis) and frequency response.
The final three splits are by time throughout the mission.
\begin{itemize}
  \item \textbf{Inner/Outer Focal Plane Rows:} Split by physical detector row on the focal plane.  This split also divides detectors according to their location in the multiplexed readout.  We have observed RF-coupled interference (including the reaction-wheel synchronous noise) that is more prevalent in the inner focal plane rows, making this split sensitive to any related residuals.
  \item \textbf{Inner/Outer Focal Plane Radius:} Split by detector distance from the center of the focal plane.
  This probes beam shape, which becomes increasingly elliptical toward the focal plane edges.
  \item \textbf{Diagonal Tiles:} Split focal planes into sets of two tiles located diagonally across from each other.
  Since each tile is fabricated independently, this tests for detector non-uniformity in fabrication.
  \item \textbf{Checkerboard:} Split the square grid of detectors in a checkerboard pattern (not splitting orthogonally polarized pairs).
  This test probes the noise model, since we do not expect any instrumental systematics to vary on this basis.
  \item \textbf{Alternating Mux Column:} Split every other readout column.
    This probes differences in the detector bias and SQUID readout among columns.
  \item \textbf{Port/Starboard Detectors:} Split by detector pointing azimuth to difference detectors located on the port and starboard side of a given receiver.
  This is sensitive to sidelobe pickup from the Galaxy and Sun, which are on opposite sides of the azimuthal scan.
  \item \textbf{Band Center:} Split all detectors at a given observing frequency by their measured band center.
  The sets have mean band center differences of 2 and \SI{4}{\giga\hertz} at observing frequencies 95 and \SI{150}{\giga\hertz}, respectively.
  This probes \spider's sensitivity to differential responsivity to Galactic dust between orthogonally polarized detectors, which could bias the foreground and cosmological results.
  \item \textbf{Left/Right Scan:} Split azimuthal scans into left-going and right-going.
  This probes time constant effects.
  \item \textbf{Alternating Days:} Split into every other day.
  This probes HWP systematics, since full polarization angle coverage for a given detector requires four independent HWP angles, which corresponds to two days of observing time.
  \item \textbf{Early/Late Flight:} Split each of the two scan strategies into early and late halves. This probes longer trends, such as effects from cryogen loss.
\end{itemize}

\subsubsection{Processing}
For each half of a null split, a data map is made with the standard processing pipeline described in Section~\ref{sec:processing}.
A simulated re-observed \planck map is also made for each \spider half-data map, using the same detector/time split.
The \planck maps provide estimates of the expected null signal residual, since they capture both CMB and foreground power; the latter is consistent with being the dominant source of signal residual power at large scales in the \spider null maps.
For most null splits, the foreground and CMB signal residuals are small compared to the noise.
However, they are significant in particular for the Port/Starboard \SI{150}{\giga\hertz} null split, in which the lowest bin's residual null power is reduced by a factor of two when accounting for foregrounds.
To perform a null test, we first subtract the simulated \planck maps from the \spider half-data maps, and then difference the two halves to form a null map.
The power spectra and covariances are estimated from the null maps with the XFaster and NSI pipelines.

XFaster uses 500 signal and noise simulations per null split, unlike the 1000 used for signal power spectra.
Additionally, the mode-loss factor, $g_\ell$, is determined differently for null spectra.
It is computed for each null using signal and noise simulations, rather than signal-only simulations, as described in Section~\ref{subsection:XFaster}.
This procedure has been validated with simulations.
NSI null spectra are computed in the same way as signal power spectra, as described in Section~\ref{subsection:NSI}.

\subsubsection{Null Test Results}
\label{subsec:null_results}
\begin{figure*}
\centering
\includegraphics[scale=1]{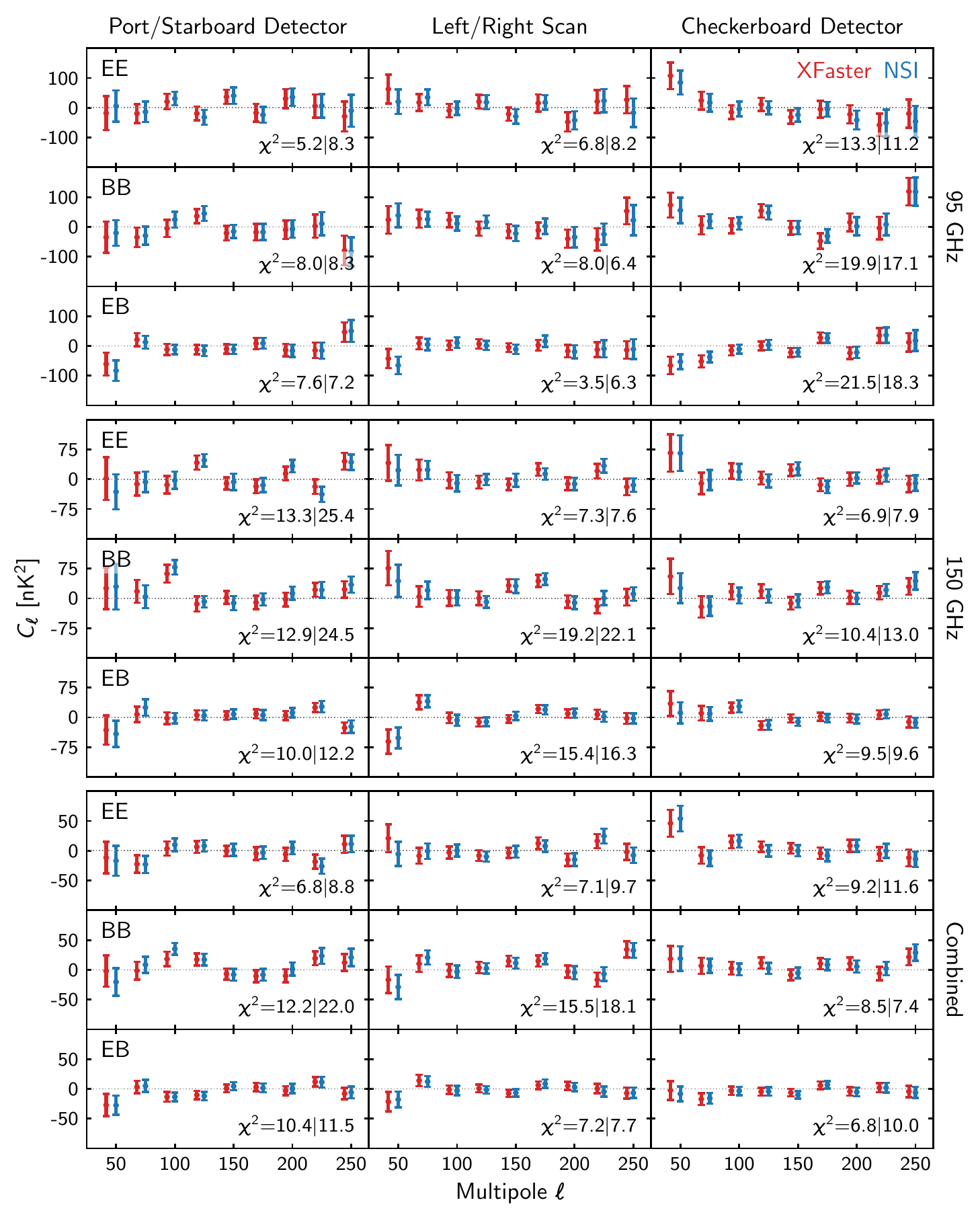}
\caption{
    Three null tests, showing the comparison between the XFaster and NSI pipelines for the different frequency combinations.
    A horizontal offset is added to NSI points for visual clarity.
    $\chi^2$ values are computed independently for each pipeline and spectrum.
    See Table~\ref{table:null_pte} for the corresponding probability-to-exceed (PTE) value for each test.}
\label{fig:null_spectra}
\end{figure*}

Both pipelines are used to construct $EE$, $BB$, and $EB$ spectra.
$\chi^2$ values are computed for each test over the nine $\ell$-bins and expressed as probability-to-exceed (PTE) values in Table~\ref{table:null_pte}.
Example null spectra are shown in Figure~\ref{fig:null_spectra}.

\begin{table}
  \centering
  \label{table:null_pte}
  \caption{Null test results expressed as probability-to-exceed (PTE) values, assuming a $\chi^2$ distribution for each.
    Results are shown for each polarization spectrum and pipeline, as well as for each frequency band and for the combined data set (``Comb'').
    While these PTE values are useful for evaluating individual tests, correlations affect their interpretation across multiple tests (see Section~\ref{subsec:null_results} and Table~\ref{table:null_stats}).
    The color scale extends from 0 (\textit{dark}) to 1 (\textit{light}) to draw attention to more unlikely PTEs at either end of the range.}
  
\newlength{\ColWidthNormal} \setlength{\ColWidthNormal}{0.5cm}
\newlength{\ColWidthSpec} \setlength{\ColWidthSpec}{0.5cm}
\newcolumntype{R}{>{\raggedleft\arraybackslash}p{\ColWidthNormal}}
\newcolumntype{L}{>{\raggedright\arraybackslash}p{\ColWidthSpec}}
\sisetup{table-format=1.3}
\begin{tabularx}{\columnwidth}{@{}L@{}R@{}R@{}R@{}R@{}R@{}R}
  \toprule
  \toprule
  & \multicolumn{3}{c|}{XFaster} & \multicolumn{3}{c}{NSI}\\
  & \multicolumn{1}{c}{\makebox[0.115\columnwidth]{\SI{95}{\giga\hertz}}} & \multicolumn{1}{c}{\makebox[0.115\columnwidth]{\SI{150}{\giga\hertz}}} & \multicolumn{1}{c|}{\makebox[0.115\columnwidth]{Comb}} & \multicolumn{1}{c}{\makebox[0.115\columnwidth]{\SI{95}{\giga\hertz}}} & \multicolumn{1}{c}{\makebox[0.115\columnwidth]{\SI{150}{\giga\hertz}}} & \multicolumn{1}{c}{\makebox[0.115\columnwidth]{Comb}}\\
  \midrule
  \multicolumn{7}{l}{Inner/Outer Focal Plane Rows} \\
  EE & \multicolumn{1}{>{\columncolor[rgb]{0.985359477124,0.457516339869,0.332026143791}[5pt]}S}{0.06} & \multicolumn{1}{>{\columncolor[rgb]{0.98891195694,0.74585159554,0.647950788158}[5pt]}S}{0.52} & \multicolumn{1}{>{\columncolor[rgb]{0.98646674356,0.501806997309,0.37631680123}[5pt]}S}{0.14} & \multicolumn{1}{>{\columncolor[rgb]{0.567058823529,0.763306420607,0.871018838908}[5pt]}S}{0.19} & \multicolumn{1}{>{\columncolor[rgb]{0.787543252595,0.866205305652,0.940945790081}[5pt]}S}{0.53} & \multicolumn{1}{>{\columncolor[rgb]{0.498039215686,0.725413302576,0.856132256824}[5pt]}S}{0.10} \\
  BB & \multicolumn{1}{>{\columncolor[rgb]{0.984375240292,0.41814686659,0.292656670511}[5pt]}S}{0.007} & \multicolumn{1}{>{\columncolor[rgb]{0.986589773164,0.506728181469,0.38123798539}[5pt]}S}{0.14} & \multicolumn{1}{>{\columncolor[rgb]{0.98585159554,0.477201076509,0.351710880431}[5pt]}S}{0.10} & \multicolumn{1}{>{\columncolor[rgb]{0.485490196078,0.718523644752,0.853425605536}[5pt]}S}{0.08} & \multicolumn{1}{>{\columncolor[rgb]{0.460392156863,0.704744329104,0.84801230296}[5pt]}S}{0.05} & \multicolumn{1}{>{\columncolor[rgb]{0.435294117647,0.690965013456,0.842599000384}[5pt]}S}{0.02} \\
  EB & \multicolumn{1}{>{\columncolor[rgb]{0.988235294118,0.676585928489,0.566320645905}[5pt]}S}{0.41} & \multicolumn{1}{>{\columncolor[rgb]{0.996432141484,0.88585928489,0.83414071511}[5pt]}S}{0.77} & \multicolumn{1}{>{\columncolor[rgb]{0.997785467128,0.914279123414,0.874740484429}[5pt]}S}{0.85} & \multicolumn{1}{>{\columncolor[rgb]{0.811164936563,0.881953094963,0.948819684737}[5pt]}S}{0.59} & \multicolumn{1}{>{\columncolor[rgb]{0.888658208381,0.933133410227,0.974409842368}[5pt]}S}{0.79} & \multicolumn{1}{>{\columncolor[rgb]{0.861361014994,0.915417147251,0.96555171088}[5pt]}S}{0.73} \\
  \multicolumn{7}{l}{Inner/Outer Focal Plane Radius} \\
  EE & \multicolumn{1}{>{\columncolor[rgb]{0.993587081892,0.832341407151,0.762491349481}[5pt]}S}{0.67} & \multicolumn{1}{>{\columncolor[rgb]{0.997908496732,0.916862745098,0.878431372549}[5pt]}S}{0.86} & \multicolumn{1}{>{\columncolor[rgb]{0.999753940792,0.955617070358,0.933794694348}[5pt]}S}{0.98} & \multicolumn{1}{>{\columncolor[rgb]{0.834786620531,0.897700884275,0.956693579393}[5pt]}S}{0.65} & \multicolumn{1}{>{\columncolor[rgb]{0.947097270281,0.970534409842,0.993110342176}[5pt]}S}{0.94} & \multicolumn{1}{>{\columncolor[rgb]{0.96555171088,0.982345251826,0.999015763168}[5pt]}S}{0.99} \\
  BB & \multicolumn{1}{>{\columncolor[rgb]{0.9846212995,0.42798923491,0.302499038831}[5pt]}S}{0.02} & \multicolumn{1}{>{\columncolor[rgb]{0.99876970396,0.934948096886,0.904267589389}[5pt]}S}{0.92} & \multicolumn{1}{>{\columncolor[rgb]{0.988235294118,0.676585928489,0.566320645905}[5pt]}S}{0.41} & \multicolumn{1}{>{\columncolor[rgb]{0.50431372549,0.728858131488,0.857485582468}[5pt]}S}{0.11} & \multicolumn{1}{>{\columncolor[rgb]{0.642368319877,0.801830065359,0.890319108035}[5pt]}S}{0.29} & \multicolumn{1}{>{\columncolor[rgb]{0.604705882353,0.783975394079,0.879138792772}[5pt]}S}{0.23} \\
  EB & \multicolumn{1}{>{\columncolor[rgb]{0.986589773164,0.506728181469,0.38123798539}[5pt]}S}{0.15} & \multicolumn{1}{>{\columncolor[rgb]{0.988235294118,0.636232218378,0.520061514802}[5pt]}S}{0.34} & \multicolumn{1}{>{\columncolor[rgb]{0.997170319108,0.901361014994,0.856286043829}[5pt]}S}{0.82} & \multicolumn{1}{>{\columncolor[rgb]{0.671895424837,0.814379084967,0.900653594771}[5pt]}S}{0.33} & \multicolumn{1}{>{\columncolor[rgb]{0.516862745098,0.735747789312,0.860192233756}[5pt]}S}{0.12} & \multicolumn{1}{>{\columncolor[rgb]{0.953248750481,0.97447135717,0.99507881584}[5pt]}S}{0.95} \\
  \multicolumn{7}{l}{Diagonal Tiles} \\
  EE & \multicolumn{1}{>{\columncolor[rgb]{0.988235294118,0.651364859669,0.537408688966}[5pt]}S}{0.37} & \multicolumn{1}{>{\columncolor[rgb]{0.986835832372,0.516570549789,0.39108035371}[5pt]}S}{0.16} & \multicolumn{1}{>{\columncolor[rgb]{0.98708189158,0.526412918108,0.40092272203}[5pt]}S}{0.18} & \multicolumn{1}{>{\columncolor[rgb]{0.681737793156,0.818562091503,0.904098423683}[5pt]}S}{0.34} & \multicolumn{1}{>{\columncolor[rgb]{0.498039215686,0.725413302576,0.856132256824}[5pt]}S}{0.10} & \multicolumn{1}{>{\columncolor[rgb]{0.472941176471,0.711633986928,0.850718954248}[5pt]}S}{0.07} \\
  BB & \multicolumn{1}{>{\columncolor[rgb]{0.9969242599,0.896193771626,0.848904267589}[5pt]}S}{0.80} & \multicolumn{1}{>{\columncolor[rgb]{0.998400615148,0.927197231834,0.893194925029}[5pt]}S}{0.90} & \multicolumn{1}{>{\columncolor[rgb]{0.997293348712,0.903944636678,0.859976931949}[5pt]}S}{0.82} & \multicolumn{1}{>{\columncolor[rgb]{0.642368319877,0.801830065359,0.890319108035}[5pt]}S}{0.28} & \multicolumn{1}{>{\columncolor[rgb]{0.787543252595,0.866205305652,0.940945790081}[5pt]}S}{0.53} & \multicolumn{1}{>{\columncolor[rgb]{0.579607843137,0.770196078431,0.873725490196}[5pt]}S}{0.20} \\
  EB & \multicolumn{1}{>{\columncolor[rgb]{0.985113417916,0.447673971549,0.322183775471}[5pt]}S}{0.05} & \multicolumn{1}{>{\columncolor[rgb]{0.997293348712,0.903944636678,0.859976931949}[5pt]}S}{0.83} & \multicolumn{1}{>{\columncolor[rgb]{0.988235294118,0.62614379085,0.508496732026}[5pt]}S}{0.33} & \multicolumn{1}{>{\columncolor[rgb]{0.491764705882,0.721968473664,0.85477893118}[5pt]}S}{0.09} & \multicolumn{1}{>{\columncolor[rgb]{0.637447135717,0.799738562092,0.888596693579}[5pt]}S}{0.28} & \multicolumn{1}{>{\columncolor[rgb]{0.541960784314,0.74952710496,0.865605536332}[5pt]}S}{0.15} \\
  \multicolumn{7}{l}{Checkerboard Detectors} \\
  EE & \multicolumn{1}{>{\columncolor[rgb]{0.986712802768,0.511649365629,0.38615916955}[5pt]}S}{0.15} & \multicolumn{1}{>{\columncolor[rgb]{0.992848904268,0.818685121107,0.744405997693}[5pt]}S}{0.64} & \multicolumn{1}{>{\columncolor[rgb]{0.988235294118,0.681630142253,0.572103037293}[5pt]}S}{0.42} & \multicolumn{1}{>{\columncolor[rgb]{0.627604767397,0.795555555556,0.885151864667}[5pt]}S}{0.26} & \multicolumn{1}{>{\columncolor[rgb]{0.793448673587,0.87014225298,0.942914263745}[5pt]}S}{0.54} & \multicolumn{1}{>{\columncolor[rgb]{0.610980392157,0.787420222991,0.880492118416}[5pt]}S}{0.24} \\
  BB & \multicolumn{1}{>{\columncolor[rgb]{0.9846212995,0.42798923491,0.302499038831}[5pt]}S}{0.02} & \multicolumn{1}{>{\columncolor[rgb]{0.988235294118,0.616055363322,0.49693194925}[5pt]}S}{0.32} & \multicolumn{1}{>{\columncolor[rgb]{0.988235294118,0.727028066128,0.624144559785}[5pt]}S}{0.49} & \multicolumn{1}{>{\columncolor[rgb]{0.460392156863,0.704744329104,0.84801230296}[5pt]}S}{0.05} & \multicolumn{1}{>{\columncolor[rgb]{0.548235294118,0.752971933872,0.866958861976}[5pt]}S}{0.16} & \multicolumn{1}{>{\columncolor[rgb]{0.814117647059,0.883921568627,0.949803921569}[5pt]}S}{0.60} \\
  EB & \multicolumn{1}{>{\columncolor[rgb]{0.984498269896,0.42306805075,0.297577854671}[5pt]}S}{0.01} & \multicolumn{1}{>{\columncolor[rgb]{0.988235294118,0.661453287197,0.548973471742}[5pt]}S}{0.39} & \multicolumn{1}{>{\columncolor[rgb]{0.993341022684,0.827789311803,0.756462898885}[5pt]}S}{0.66} & \multicolumn{1}{>{\columncolor[rgb]{0.447843137255,0.69785467128,0.845305651672}[5pt]}S}{0.03} & \multicolumn{1}{>{\columncolor[rgb]{0.701422529796,0.826928104575,0.910988081507}[5pt]}S}{0.38} & \multicolumn{1}{>{\columncolor[rgb]{0.681737793156,0.818562091503,0.904098423683}[5pt]}S}{0.35} \\
  \multicolumn{7}{l}{Alternating Mux Columns} \\
  EE & \multicolumn{1}{>{\columncolor[rgb]{0.992110726644,0.805028835063,0.726320645905}[5pt]}S}{0.62} & \multicolumn{1}{>{\columncolor[rgb]{0.98523644752,0.452595155709,0.327104959631}[5pt]}S}{0.06} & \multicolumn{1}{>{\columncolor[rgb]{0.991618608228,0.795924644368,0.714263744714}[5pt]}S}{0.61} & \multicolumn{1}{>{\columncolor[rgb]{0.870219146482,0.921322568243,0.968504421376}[5pt]}S}{0.75} & \multicolumn{1}{>{\columncolor[rgb]{0.554509803922,0.756416762784,0.86831218762}[5pt]}S}{0.17} & \multicolumn{1}{>{\columncolor[rgb]{0.716186082276,0.833202614379,0.916155324875}[5pt]}S}{0.40} \\
  BB & \multicolumn{1}{>{\columncolor[rgb]{0.988235294118,0.62614379085,0.508496732026}[5pt]}S}{0.33} & \multicolumn{1}{>{\columncolor[rgb]{0.988235294118,0.621099577086,0.502714340638}[5pt]}S}{0.33} & \multicolumn{1}{>{\columncolor[rgb]{0.984867358708,0.43783160323,0.312341407151}[5pt]}S}{0.03} & \multicolumn{1}{>{\columncolor[rgb]{0.937870049981,0.96462898885,0.99015763168}[5pt]}S}{0.92} & \multicolumn{1}{>{\columncolor[rgb]{0.523137254902,0.739192618224,0.8615455594}[5pt]}S}{0.13} & \multicolumn{1}{>{\columncolor[rgb]{0.516862745098,0.735747789312,0.860192233756}[5pt]}S}{0.12} \\
  EB & \multicolumn{1}{>{\columncolor[rgb]{0.988235294118,0.721983852364,0.618362168397}[5pt]}S}{0.48} & \multicolumn{1}{>{\columncolor[rgb]{0.992110726644,0.805028835063,0.726320645905}[5pt]}S}{0.62} & \multicolumn{1}{>{\columncolor[rgb]{0.995309496348,0.864206074587,0.804690503652}[5pt]}S}{0.72} & \multicolumn{1}{>{\columncolor[rgb]{0.701422529796,0.826928104575,0.910988081507}[5pt]}S}{0.38} & \multicolumn{1}{>{\columncolor[rgb]{0.730949634756,0.839477124183,0.921322568243}[5pt]}S}{0.43} & \multicolumn{1}{>{\columncolor[rgb]{0.840692041522,0.901637831603,0.958662053057}[5pt]}S}{0.66} \\
  \multicolumn{7}{l}{Port/Starboard Pointing Detectors} \\
  EE & \multicolumn{1}{>{\columncolor[rgb]{0.997170319108,0.901361014994,0.856286043829}[5pt]}S}{0.82} & \multicolumn{1}{>{\columncolor[rgb]{0.986712802768,0.511649365629,0.38615916955}[5pt]}S}{0.15} & \multicolumn{1}{>{\columncolor[rgb]{0.993341022684,0.827789311803,0.756462898885}[5pt]}S}{0.66} & \multicolumn{1}{>{\columncolor[rgb]{0.778685121107,0.86029988466,0.937993079585}[5pt]}S}{0.51} & \multicolumn{1}{>{\columncolor[rgb]{0.422745098039,0.684075355632,0.839892349097}[5pt]}S}{0.003} & \multicolumn{1}{>{\columncolor[rgb]{0.750634371396,0.847843137255,0.928212226067}[5pt]}S}{0.46} \\
  BB & \multicolumn{1}{>{\columncolor[rgb]{0.989158016148,0.750403690888,0.653979238754}[5pt]}S}{0.53} & \multicolumn{1}{>{\columncolor[rgb]{0.986958861976,0.521491733948,0.39600153787}[5pt]}S}{0.17} & \multicolumn{1}{>{\columncolor[rgb]{0.987574009996,0.546097654748,0.42060745867}[5pt]}S}{0.20} & \multicolumn{1}{>{\columncolor[rgb]{0.778685121107,0.86029988466,0.937993079585}[5pt]}S}{0.50} & \multicolumn{1}{>{\columncolor[rgb]{0.422745098039,0.684075355632,0.839892349097}[5pt]}S}{0.004} & \multicolumn{1}{>{\columncolor[rgb]{0.429019607843,0.687520184544,0.84124567474}[5pt]}S}{0.009} \\
  EB & \multicolumn{1}{>{\columncolor[rgb]{0.990634371396,0.777716262976,0.69014994233}[5pt]}S}{0.57} & \multicolumn{1}{>{\columncolor[rgb]{0.988235294118,0.636232218378,0.520061514802}[5pt]}S}{0.35} & \multicolumn{1}{>{\columncolor[rgb]{0.988235294118,0.621099577086,0.502714340638}[5pt]}S}{0.32} & \multicolumn{1}{>{\columncolor[rgb]{0.820023068051,0.887858515955,0.951772395233}[5pt]}S}{0.61} & \multicolumn{1}{>{\columncolor[rgb]{0.585882352941,0.773640907343,0.87507881584}[5pt]}S}{0.20} & \multicolumn{1}{>{\columncolor[rgb]{0.617254901961,0.790865051903,0.88184544406}[5pt]}S}{0.25} \\
  \multicolumn{7}{l}{Band Center} \\
  EE & \multicolumn{1}{>{\columncolor[rgb]{0.99630911188,0.883275663206,0.83044982699}[5pt]}S}{0.76} & \multicolumn{1}{>{\columncolor[rgb]{0.988235294118,0.666497500961,0.55475586313}[5pt]}S}{0.39} & \multicolumn{1}{>{\columncolor[rgb]{0.998892733564,0.93753171857,0.907958477509}[5pt]}S}{0.92} & \multicolumn{1}{>{\columncolor[rgb]{0.96247597078,0.980376778162,0.998031526336}[5pt]}S}{0.98} & \multicolumn{1}{>{\columncolor[rgb]{0.617254901961,0.790865051903,0.88184544406}[5pt]}S}{0.24} & \multicolumn{1}{>{\columncolor[rgb]{0.944021530181,0.968565936178,0.992126105344}[5pt]}S}{0.93} \\
  BB & \multicolumn{1}{>{\columncolor[rgb]{0.987327950788,0.536255286428,0.41076509035}[5pt]}S}{0.19} & \multicolumn{1}{>{\columncolor[rgb]{0.987204921184,0.531334102268,0.40584390619}[5pt]}S}{0.18} & \multicolumn{1}{>{\columncolor[rgb]{0.99815455594,0.922029988466,0.885813148789}[5pt]}S}{0.88} & \multicolumn{1}{>{\columncolor[rgb]{0.604705882353,0.783975394079,0.879138792772}[5pt]}S}{0.23} & \multicolumn{1}{>{\columncolor[rgb]{0.548235294118,0.752971933872,0.866958861976}[5pt]}S}{0.16} & \multicolumn{1}{>{\columncolor[rgb]{0.726028450596,0.837385620915,0.919600153787}[5pt]}S}{0.42} \\
  EB & \multicolumn{1}{>{\columncolor[rgb]{0.988665897732,0.741299500192,0.641922337562}[5pt]}S}{0.51} & \multicolumn{1}{>{\columncolor[rgb]{0.995309496348,0.864206074587,0.804690503652}[5pt]}S}{0.72} & \multicolumn{1}{>{\columncolor[rgb]{0.995555555556,0.868758169935,0.810718954248}[5pt]}S}{0.73} & \multicolumn{1}{>{\columncolor[rgb]{0.843644752018,0.903606305267,0.959646289889}[5pt]}S}{0.67} & \multicolumn{1}{>{\columncolor[rgb]{0.822975778547,0.889826989619,0.952756632065}[5pt]}S}{0.62} & \multicolumn{1}{>{\columncolor[rgb]{0.820023068051,0.887858515955,0.951772395233}[5pt]}S}{0.61} \\
  \multicolumn{7}{l}{Left/Right Scan} \\
  EE & \multicolumn{1}{>{\columncolor[rgb]{0.993341022684,0.827789311803,0.756462898885}[5pt]}S}{0.66} & \multicolumn{1}{>{\columncolor[rgb]{0.991618608228,0.795924644368,0.714263744714}[5pt]}S}{0.61} & \multicolumn{1}{>{\columncolor[rgb]{0.992110726644,0.805028835063,0.726320645905}[5pt]}S}{0.62} & \multicolumn{1}{>{\columncolor[rgb]{0.781637831603,0.862268358324,0.938977316417}[5pt]}S}{0.51} & \multicolumn{1}{>{\columncolor[rgb]{0.805259515571,0.878016147636,0.946851211073}[5pt]}S}{0.57} & \multicolumn{1}{>{\columncolor[rgb]{0.696501345636,0.824836601307,0.909265667051}[5pt]}S}{0.37} \\
  BB & \multicolumn{1}{>{\columncolor[rgb]{0.989404075356,0.754955786236,0.66000768935}[5pt]}S}{0.54} & \multicolumn{1}{>{\columncolor[rgb]{0.9846212995,0.42798923491,0.302499038831}[5pt]}S}{0.02} & \multicolumn{1}{>{\columncolor[rgb]{0.985605536332,0.467358708189,0.341868512111}[5pt]}S}{0.08} & \multicolumn{1}{>{\columncolor[rgb]{0.852502883506,0.909511726259,0.962599000384}[5pt]}S}{0.70} & \multicolumn{1}{>{\columncolor[rgb]{0.429019607843,0.687520184544,0.84124567474}[5pt]}S}{0.009} & \multicolumn{1}{>{\columncolor[rgb]{0.447843137255,0.69785467128,0.845305651672}[5pt]}S}{0.03} \\
  EB & \multicolumn{1}{>{\columncolor[rgb]{0.999138792772,0.942698961938,0.915340253749}[5pt]}S}{0.94} & \multicolumn{1}{>{\columncolor[rgb]{0.985605536332,0.467358708189,0.341868512111}[5pt]}S}{0.08} & \multicolumn{1}{>{\columncolor[rgb]{0.991864667436,0.800476739715,0.720292195309}[5pt]}S}{0.62} & \multicolumn{1}{>{\columncolor[rgb]{0.858408304498,0.913448673587,0.964567474048}[5pt]}S}{0.71} & \multicolumn{1}{>{\columncolor[rgb]{0.466666666667,0.708189158016,0.849365628604}[5pt]}S}{0.06} & \multicolumn{1}{>{\columncolor[rgb]{0.802306805075,0.876047673972,0.945866974241}[5pt]}S}{0.57} \\
  \multicolumn{7}{l}{Alternating Days} \\
  EE & \multicolumn{1}{>{\columncolor[rgb]{0.984990388312,0.442752787389,0.317262591311}[5pt]}S}{0.04} & \multicolumn{1}{>{\columncolor[rgb]{0.988066128412,0.565782391388,0.440292195309}[5pt]}S}{0.24} & \multicolumn{1}{>{\columncolor[rgb]{0.994079200308,0.841445597847,0.774548250673}[5pt]}S}{0.68} & \multicolumn{1}{>{\columncolor[rgb]{0.472941176471,0.711633986928,0.850718954248}[5pt]}S}{0.07} & \multicolumn{1}{>{\columncolor[rgb]{0.604705882353,0.783975394079,0.879138792772}[5pt]}S}{0.23} & \multicolumn{1}{>{\columncolor[rgb]{0.904036908881,0.942975778547,0.979331026528}[5pt]}S}{0.83} \\
  BB & \multicolumn{1}{>{\columncolor[rgb]{0.984867358708,0.43783160323,0.312341407151}[5pt]}S}{0.04} & \multicolumn{1}{>{\columncolor[rgb]{0.993341022684,0.827789311803,0.756462898885}[5pt]}S}{0.66} & \multicolumn{1}{>{\columncolor[rgb]{0.988235294118,0.580745866974,0.456455209535}[5pt]}S}{0.27} & \multicolumn{1}{>{\columncolor[rgb]{0.610980392157,0.787420222991,0.880492118416}[5pt]}S}{0.24} & \multicolumn{1}{>{\columncolor[rgb]{0.814117647059,0.883921568627,0.949803921569}[5pt]}S}{0.60} & \multicolumn{1}{>{\columncolor[rgb]{0.691580161476,0.822745098039,0.907543252595}[5pt]}S}{0.37} \\
  EB & \multicolumn{1}{>{\columncolor[rgb]{0.999138792772,0.942698961938,0.915340253749}[5pt]}S}{0.94} & \multicolumn{1}{>{\columncolor[rgb]{0.985113417916,0.447673971549,0.322183775471}[5pt]}S}{0.05} & \multicolumn{1}{>{\columncolor[rgb]{0.997416378316,0.906528258362,0.863667820069}[5pt]}S}{0.83} & \multicolumn{1}{>{\columncolor[rgb]{0.760476739715,0.852026143791,0.931657054979}[5pt]}S}{0.47} & \multicolumn{1}{>{\columncolor[rgb]{0.610980392157,0.787420222991,0.880492118416}[5pt]}S}{0.24} & \multicolumn{1}{>{\columncolor[rgb]{0.937870049981,0.96462898885,0.99015763168}[5pt]}S}{0.92} \\
  \multicolumn{7}{l}{Early/Late Flight} \\
  EE & \multicolumn{1}{>{\columncolor[rgb]{0.996555171088,0.888442906574,0.83783160323}[5pt]}S}{0.78} & \multicolumn{1}{>{\columncolor[rgb]{0.984498269896,0.42306805075,0.297577854671}[5pt]}S}{0.008} & \multicolumn{1}{>{\columncolor[rgb]{0.994325259516,0.845997693195,0.780576701269}[5pt]}S}{0.69} & \multicolumn{1}{>{\columncolor[rgb]{0.86431372549,0.917385620915,0.966535947712}[5pt]}S}{0.73} & \multicolumn{1}{>{\columncolor[rgb]{0.53568627451,0.746082276048,0.864252210688}[5pt]}S}{0.15} & \multicolumn{1}{>{\columncolor[rgb]{0.922491349481,0.954786620531,0.98523644752}[5pt]}S}{0.88} \\
  BB & \multicolumn{1}{>{\columncolor[rgb]{0.985605536332,0.467358708189,0.341868512111}[5pt]}S}{0.08} & \multicolumn{1}{>{\columncolor[rgb]{0.98646674356,0.501806997309,0.37631680123}[5pt]}S}{0.14} & \multicolumn{1}{>{\columncolor[rgb]{0.986589773164,0.506728181469,0.38123798539}[5pt]}S}{0.14} & \multicolumn{1}{>{\columncolor[rgb]{0.647289504037,0.803921568627,0.892041522491}[5pt]}S}{0.29} & \multicolumn{1}{>{\columncolor[rgb]{0.676816608997,0.816470588235,0.902376009227}[5pt]}S}{0.34} & \multicolumn{1}{>{\columncolor[rgb]{0.775240292195,0.858300653595,0.936824298347}[5pt]}S}{0.50} \\
  EB & \multicolumn{1}{>{\columncolor[rgb]{0.985359477124,0.457516339869,0.332026143791}[5pt]}S}{0.07} & \multicolumn{1}{>{\columncolor[rgb]{0.988235294118,0.691718569781,0.583667820069}[5pt]}S}{0.43} & \multicolumn{1}{>{\columncolor[rgb]{0.988235294118,0.686674356017,0.577885428681}[5pt]}S}{0.42} & \multicolumn{1}{>{\columncolor[rgb]{0.466666666667,0.708189158016,0.849365628604}[5pt]}S}{0.06} & \multicolumn{1}{>{\columncolor[rgb]{0.447843137255,0.69785467128,0.845305651672}[5pt]}S}{0.04} & \multicolumn{1}{>{\columncolor[rgb]{0.585882352941,0.773640907343,0.87507881584}[5pt]}S}{0.21} \\
  \bottomrule
\end{tabularx}

\end{table}

\begin{table}
  \centering
  \label{table:null_stats}
    \caption{Null test ensemble results, given as the proportion of simulations with worse test statistics than the data.
    The outlier test probes the number of simulations with at least as large of a $\chi^2$ as the largest data $\chi^2$.
    The distribution test probes the shape of the null statistics across all tests, accounting for correlations among similar null splits.
    For the XFaster pipeline, this is computed as a KS-test using simulations to calibrate the bias due to correlations, and for NSI as a single ``combined $\chi^2$''.}
    \newcolumntype{C}{>{\centering\arraybackslash}p{0.155\columnwidth}}
    \begin{tabularx}{\columnwidth}{lCCCC}
    \toprule
    \toprule
    ~ & \multicolumn{2}{c}{Outlier Test PTE} & \multicolumn{2}{c}{Distribution Test PTE} \\
    Band & XFaster & NSI & XFaster & NSI \\
    \midrule
    \SI{95}{\giga\hertz} & 0.38 & 0.80 & 0.07 & $\text{N/A}$ \\
    \SI{150}{\giga\hertz} & 0.34 & 0.20 & 0.21 & $\text{N/A}$ \\
    Combined & 0.78 & 0.34 & 0.56 & 0.50 \\
    \bottomrule
    \end{tabularx}
\end{table}

For a large ensemble of uncorrelated, noise-dominated spectra, the PTE values are expected to be uniformly distributed between 0 and 1, with no extreme outliers at either end of the distribution.
Non-negligible correlations among the null splits are known to exist, however, because they share detector samples.
Thus, the distributions of the null spectra are evaluated using simulations incorporating these correlations.
For XFaster, null bandpowers and covariance matrices are computed from 500 simulated signal and noise maps, each seed of which naturally incorporates correlations across null splits.
For NSI, simulated null bandpowers are instead generated from realizations of the covariance matrix among the null test bandpowers, itself estimated from the ensemble of chunk cross-spectra.
Two tests are conducted on the resulting null statistic distributions: one probing the distributions' outliers, the other their shapes.
We require that each test results in a PTE for the observed data of at least \SI{1}{\percent}.

In the outlier test, we count how many simulations have a largest $\chi^2$ at least as high as the largest $\chi^2$ measured for the data.
The results are shown in Table~\ref{table:null_stats} in the Outlier Test column, with both pipelines passing this test.

The NSI distribution shape test is computed as a single ``combined $\chi^2$'' from a single covariance matrix including all of the null tests.
The resulting value is compared to simulations drawn from the combined covariance matrix, with a $p$-value calculated as the fraction of simulations with a higher combined $\chi^2$ than the data.
Because the combined covariance matrix is 270$\times$270 (270=10 splits $\times$ 3 spectra $\times$ 9 bins), only the combined frequency case with 378 cross-spectra is sufficient to compute it.
Thus, the individual frequency $\chi^2$ distribution tests rely on the XFaster result.

The XFaster distribution shape test is performed using Kolmogorov--Smirnov (KS) tests.
A KS-test $p$-value for the data is computed by comparing the 30 data $\chi^2$ values (10 null splits $\times$ 3 spectra) to the distribution of 150,000 simulated $\chi^2$ values (5000 draws $\times$ 10 null splits $\times$ 3 spectra) from random bandpower draws from the data covariance matrix.
This exercise is then repeated for each of the 500 simulations.
This gives a $p$-value per simulation, each of which intrinsically includes the effect of correlations between null splits.
We then determine the number of simulations with a lower $p$-value than the data, and find that all frequency combinations pass our threshold of \SI{1}{\percent}, as shown in the Distribution Test column of Table~\ref{table:null_stats}.

Based on results of these tests, data from one of the \SI{150}{\giga\hertz} receivers were dropped from the present analysis;  its data are excluded from all results shown in this paper.
This receiver was uniquely susceptible to noise correlated with the reaction wheel angle.
Because all other receivers pass the Inner/Outer Focal Plane Row null test, we have confidence that the same systematic issues do not affect the rest of the data used for this analysis.

\subsection{Raw Spectrum Comparison}
\label{subsec:est_consistency}
\begin{figure*}
  \centering
  \includegraphics[width=\textwidth]{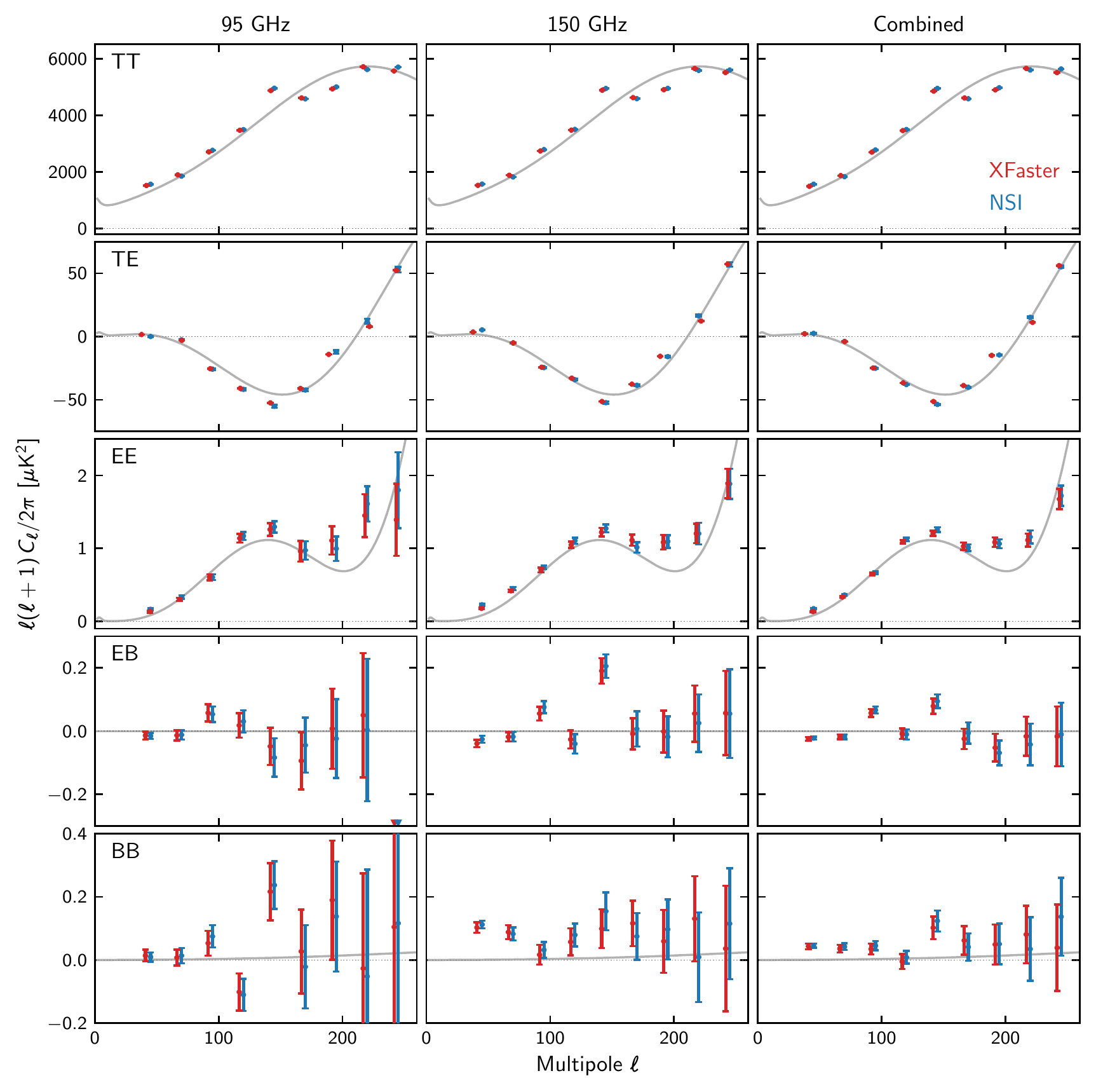}
  \caption{Raw power spectra from the XFaster and NSI pipelines.  Spectra are
    computed for each of the two frequency bands individually in the left and
    middle column, and the combined best estimate spectrum from both frequencies
    is shown in the right column.  The best fit \planck\ $\Lambda$CDM power
    spectrum from \cite{planck15_params} is shown in gray.  Error bars do not
    include sample variance, in order to better compare the instrumental noise
    estimates between the two power spectrum pipelines.  The $TB$ spectrum,
    which is not used in this cosmological analysis, is omitted.}
  \label{fig:raw_spectra}
\end{figure*}

While the null tests provide important consistency checks on the \spider data,
the two independent power spectrum estimation pipelines also provide an important
consistency check on the methodology. Figure~\ref{fig:raw_spectra} shows the power spectra from both pipelines
for the full \spider data set.
The bandpower error bars for both pipelines include only the instrumental
noise contribution (no sample variance) for ease of comparison.
Since no foreground cleaning has
been applied to these raw power spectra, excess power over the $\Lambda$CDM model is expected, particularly on large angular scales.  Small differences are observed in the bandpowers from each pipeline, particularly in the lowest multipole bins of the $TT$ and $TE$ spectra. These derive from differences in the estimators, notably effects of the different bandpower window functions and transfer functions.  These are most important in the lowest bins but are handled consistently in the likelihood analysis described in Section~\ref{sec:likelihoods}.

\section{Systematic Error Budget}
\label{sec:seb}
\begin{figure}
  \centering
  \includegraphics[width=\columnwidth]{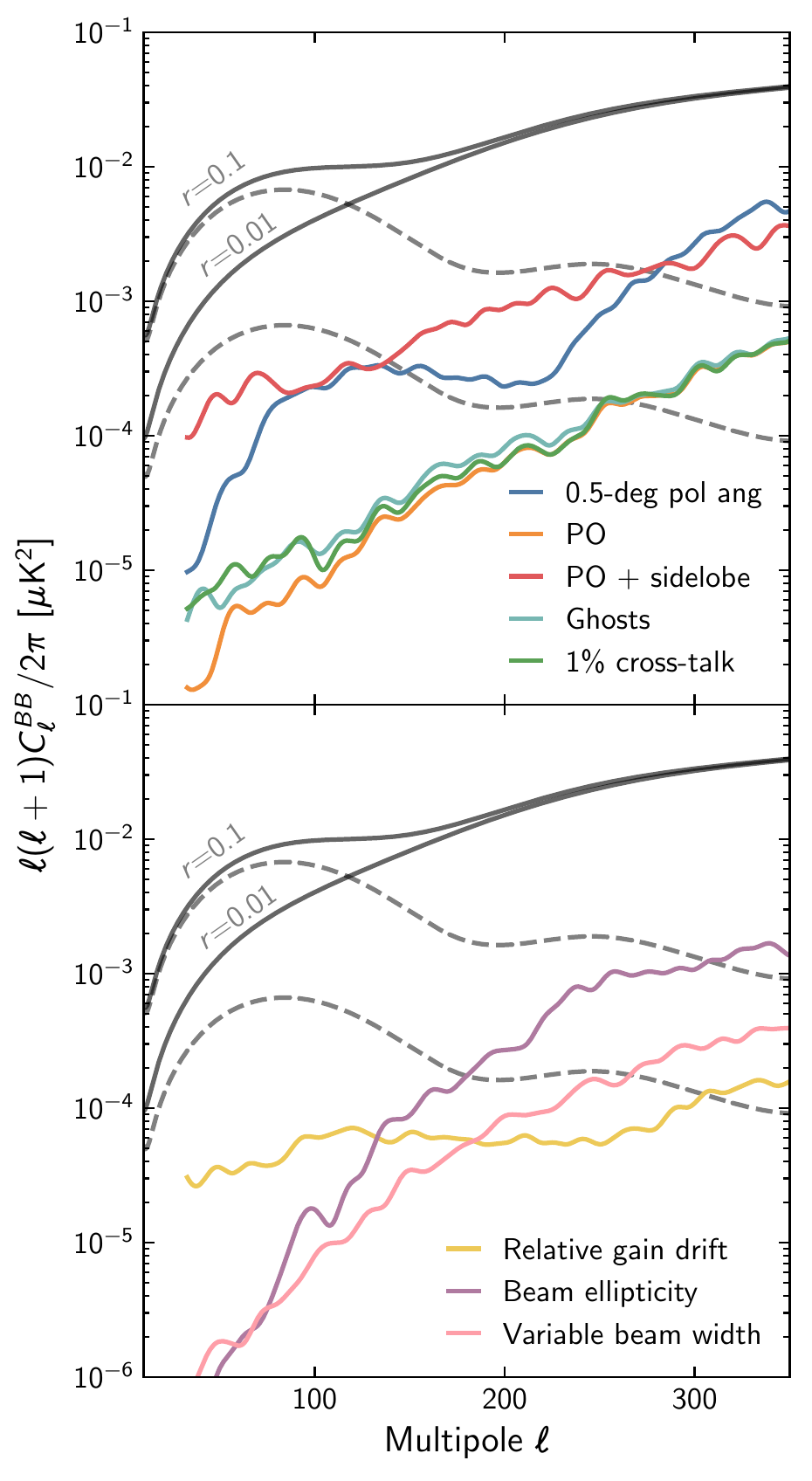}
  \caption{Simulated residual $B$-mode power from several systematic effects at \SI{150}{\giga\hertz}. The top panel shows the residuals from offset detector orientation angles and beam model extensions simulated using the \texttt{beamconv} algorithm. The bottom panel shows residuals determined from time domain deprojection templates. Legend entries are described further in the main text. Also shown for comparison are total $BB$ spectra with lensing (\textit{solid gray}) and without (\textit{dashed gray}) for two benchmark values of $r$.}
  \label{fig:bb_residuals}
\end{figure}

We use an ensemble of time-domain simulations to study the impact of various instrumental systematic effects on our ability to constrain a $B$-mode signal.
We consider eight classes of systematic effect previously identified as relevant for \spider \citep{Fraisse}, spanning optical non-idealities, calibration errors, and electrical cross-talk among detectors.
For each of these we re-observe simulated sky maps with the full filtering
and map-making pipeline (Section~\ref{sec:sims}) while injecting systematic signal into the time-ordered data.
The simulated sky maps are generated from the \Planck best-fit $\Lambda$CDM power spectrum~\citep{Planck_ref}, but with no input $B$-mode power; any output $BB$ power is thus ascribed to systematic effects. 
Figure~\ref{fig:bb_residuals} shows the impact of these effects on the $B$-mode power spectrum at \SI{150}{\giga\hertz}; the result for \SI{95}{\giga\hertz} is qualitatively similar.
Of the eight systematic effects considered, none are large enough to meaningfully impact \spider's measured $B$-mode power spectrum.  We rely on our null tests to provide limits on the contribution of known, and unknown, sources of systematic error, including those that cannot be reliably simulated.

We start by considering an offset in the detectors' polarization orientation angles, which causes $E$-mode sky signal to be misinterpreted as $B$-modes. 
We simulate this by introducing a \ang{0.5} common (shared among all detectors) offset between the polarization angles used during re-observation and map-making.
This simulated offset is taken from the per-detector error determined in pre-flight characterization (Section~\ref{sec:trpns}).
Note that a polarization error that varies among detectors would generally average down to less net effect, so the use of a common offset is conservative. 
The simulations indicate no significant contamination even for this pessimistic case.

To investigate deviations from the idealized Gaussian beam model, we consider beams derived from Physical Optics (PO) simulations of the full \Spider telescopes.
This model was developed using the GRASP software package \citep{TicraTools}, and includes variations in beam width, non-Gaussian shape, and cross-polarization response across the focal plane.
These PO beam models are truncated to an angular extent of \ang{2.5}; they therefore do not constrain the impact of extended sidelobes.
We further explore the impact of large-amplitude sidelobe response using a conservative model of the \spider beam derived from advanced optical simulations.
These beam models are convolved with simulated skies using the \texttt{beamconv} algorithm \citep{Duivenvoorden2018}, which allows efficient generation of detector timestreams for these more general beam models.
To probe far-sidelobe coupling to the Galaxy, we add \texttt{Commander} \citep{planck_foregrounds} dust and synchrotron foreground templates to the CMB signal outside the nominal \Spider observation region.

We further use \texttt{beamconv} to simulate the instrument's response to two more complex effects: reflective ghosts within the optical system (modeled as \SI{1}{\percent} of each main beam's magnitude, comparable to laboratory tests and in-flight estimates) and cross-talk between detectors (modeled as \SI{1}{\percent} between channels in adjacent multiplexing rows; lab testing is consistent with $<\SI{0.5}{\percent}$, as expected from the readout system~\cite{deKorte2003,bicep2_instrument}).
Results of these simulations are shown in the upper panel of Figure~\ref{fig:bb_residuals}; 
all have negligible impact.

In addition to the PO beam investigation, we simulate other residuals estimated as part of the time-domain analysis.
Detector gain drifts over time are injected  based on the measurements from TES resistance (Section~\ref{sec:beams_cal}).
Per-detector deviations from the simple beam model are injected with best-fit deprojection templates (Section~\ref{sec:deprojection}) for beam width and ellipticity.
These simulations inject the full estimated effects and assume no attempt to correct for them.
This assumption is conservative for gain drifts, where the estimated amplitude is corrected for in the data analysis.
Still, as shown in the lower panel of Figure~\ref{fig:bb_residuals}, these beam and gain systematics are of negligible importance.

\section{Component Separation}
\label{sec:fgs}

Measurements of CMB polarization, particularly on large angular
scales, are complicated by the need to model and remove diffuse
Galactic emission. Modeling the Galactic signal represents one of the
most challenging obstacles in characterizing the polarization of the
CMB. We implement a variety of methods to disentangle the Galactic and
cosmological signals, each subject to different assumptions,
and assess their consistency.
This paper focuses on the CMB component estimate, while detailed discussion of 
foreground components is left for future work.

In Section~\ref{sec:template_sub} we present the template-subtraction method
that serves as the nominal foreground removal technique for the cosmological
results in this paper. Section~\ref{sec:smica} describes an implementation of
the SMICA component separation method~\citep{SMICA} on the \spider data. 
These two methods---the former map-based, the latter operating in the harmonic
domain---are both used to compute foreground-cleaned spectra, and are propagated to $r$
constraints in Section~\ref{sec:likelihoods}.
Both analyses assume that the emission from
interstellar dust is the only polarized foreground in the \spider data.
We use another harmonic-space method, described
in Section~\ref{sec:a_la_Choi} and building upon the \citet{Choi+Page} analysis,
to assess the contribution of the Galactic synchrotron emission to the polarized
signal measured by \spider, finding it negligible for the present purpose.

\subsection{Template Subtraction}
\label{sec:template_sub}

Under the assumption that the spatial morphology of the polarized emission from
interstellar dust is frequency-independent, this Galactic component can be
projected out of a map at a given frequency by fitting a scalar amplitude to
a morphological template of the emission. This approach has been successfully applied in the analysis of data from \WMAP and \Planck, and studied in the context of future orbital CMB missions~\citep[see the review by \citet{Delabrouille+Cardoso_2007} as well as][]{Dunkley_2009, Katayama_2011, efstathiou2020}.

\subsubsection{Implementation}

We model the polarized intensity measured by \Spider in a given pixel as
\begin{equation}
  \label{eq:dust1}
  S_\nu = S^{\,\mathrm{CMB}} + A_{\nu,\nu_0} S_{\nu_0}^{\,\mathrm{dust}} + n_\nu\,,
\end{equation}
where $S$ is a Stokes parameter, $\nu$ is the frequency of the map, $\nu_0$ is
the frequency at which the template is defined, $A_{\nu,\nu_0}$ is a scalar amplitude, and $n_\nu$
is the map noise. We construct dust template maps, $S^{\,\mathrm{t}}_{\nu_0}$,
from the \Planck data by subtracting the \SI{100}{\giga\hertz} map from a map at higher
frequency $\nu_0$ dominated by dust emission. With the notation in
Equation~\ref{eq:dust1},

\begin{equation}
  \label{eq:dust_temp}
  S^{\,\mathrm{t}}_{\nu_0} = S_{\nu_0} - S_{100}
                           = (1-A_{100,\nu_0})\,S_{\nu_0}^{\,\mathrm{dust}}
                             + n^{\,\mathrm{t}}_{\nu_0}\,,
\end{equation}
where $n^{\,\mathrm{t}}_{\nu_0} \equiv n_{\nu_0} - n_{100}$ is the template
noise. 

A \Spider map $S_\nu$ is cleaned by subtracting from it a dust
template $S_{\nu_0}^{\,\mathrm{t}}$ multiplied by a scalar $\alpha$. With the
notation above,
\begin{equation}
  \label{eq:map_clean}
  \begin{aligned}
    S_\nu^{\,\mathrm{cleaned}} & = S_\nu - \alpha S_{\nu_0}^{\,\mathrm{t}}\,, \\
                               & = S^{\,\mathrm{CMB}} +
                                   \left(A_{\nu,\nu_0} -
                                         \alpha[1-A_{100,\nu_0}]\right)S_{\nu_0}^{\,\mathrm{dust}}\, \\
                               & \hphantom{=}\, + n_\nu - \alpha n_{\nu_0}^{\,\mathrm{t}}\,.
  \end{aligned}
\end{equation}
We then fit for $\alpha$ to minimize dust contamination.
For each \spider map frequency and choice of dust template, the NSI pipeline finds the
value of $\alpha$ that minimizes the summed power in the lowest
three multipole bins ($33 \le \ell \leq 107$) of the cleaned $EE$ spectrum.
In the XFaster pipeline we
fit $\alpha$ and $r$ simultaneously in the likelihood using all nine multipole bins of the $EE$
and $BB$ spectra.

Finally, we note that two versions of each template are constructed, each using data 
from only one \Planck half-mission.
This
allows both the NSI and XFaster pipelines to
compute the required template-subtracted spectra as cross-spectra
between two maps with independent template noise. This eliminates
the significant noise bias that would come from the noise
auto-spectrum of a full-mission dust template, albeit at the cost of
an increase in template~noise.

\subsubsection{Template-Subtraction Results}
\label{subsubsection:MapCleanResults}

\begin{table}
  \centering
  \caption{XFaster- and NSI-fitted values of the dust-template frequency scaling
  factor at \SI{95}{\giga\hertz} ($\alpha_{95}$) and \SI{150}{\giga\hertz} ($\alpha_{150}$) for
  $\nu_0=\SI{353}{\giga\hertz}$ and $\nu_0=\SI{217}{\giga\hertz}$ (Equation~\ref{eq:map_clean}). Assuming
  a modified-blackbody dust SED with temperature
  $T_\mathrm{d}=\SI{19.6}{\kelvin}$~\citep{Planck_2018_XI}, we derive from each $\alpha$
  value the dust spectral index $\beta_\mathrm{d}$, which can be compared
  directly to the SMICA measurement of this parameter and to its value derived
  by \Planck over \SI{71}{\percent} of the high-Galactic-latitude sky. For ease
  of comparison, we also report the $\alpha$ values at \spider frequencies
  expected from the dust spectral index derived by \Planck. The SMICA recovered
  value depends on the noise model at \SI{353}{\giga\hertz}. We explore two options; the
  first assumes a noise model that is the ensemble average of FFP10
  simulations. The second assumes a noise model from the difference between the
  auto-spectrum of the full mission map and the cross-spectrum of the two
  half-mission maps.}
  \label{table:alpha_meas}
  \newcommand{\asymerror}[3]{\multicolumn{1}{c}{#1\,\raisebox{0.4ex}{\tiny$^{#2}_{#3}$}}}
  \sisetup{separate-uncertainty=true}
  \begin{tabularx}{\columnwidth}{@{}lrS[table-format=3.0(2)]S[table-format=3.0(2)]S[table-format=1.2(2)]S[table-format=1.2(2)]}
   \toprule
   \toprule
   & & \multicolumn{1}{c}{$10^3\,\alpha_{95}$} & \multicolumn{1}{c}{$10^3\,\alpha_{150}$} & \multicolumn{1}{c}{$\beta_\mathrm{d}^{95}$} & \multicolumn{1}{c}{$\beta_\mathrm{d}^{150}$} \\
   \midrule
   \multicolumn{6}{l}{Template: $\nu_0 = \SI{353}{\giga\hertz}$} \\
   & \planck & \num{16.8 \pm 0.5} & \num{44.4 \pm 0.8} & \multicolumn{2}{S[table-format=1.2(2)]}{1.53 \pm 0.02} \\
   & XFaster & 18 \pm 2 & 45 \pm 2 & \asymerror{1.49}{+0.07}{-0.09} & 1.52 \pm 0.05 \\
   & NSI & 19 \pm 5 & 45 \pm 4 & \asymerror{1.44}{+0.22}{-0.17} & 1.51 \pm 0.10 \\
   \midrule
   \multicolumn{6}{l}{Template: $\nu_0 = \SI{217}{\giga\hertz}$} \\
   & \planck & 153 \pm 3 & 404 \pm 4 & \multicolumn{2}{S[table-format=1.2(2)]}{1.53 \pm 0.02} \\
   & XFaster & 159 \pm 17 & 377 \pm  16 & \asymerror{1.51}{+0.10}{-0.12} & \asymerror{1.68}{+0.08}{-0.09} \\
   & NSI & 140 \pm 50 & 350 \pm 58 & \asymerror{1.63}{+0.46}{-0.31} & \asymerror{1.81}{+0.38}{-0.31} \\
   \midrule
   \multicolumn{6}{l}{SMICA} \\
   & FFP10 & \multicolumn{1}{c}{---} & \multicolumn{1}{c}{---} & \multicolumn{2}{S[table-format=1.2(2)]}{1.43 \pm 0.04} \\
   & Auto-Cross & \multicolumn{1}{c}{---} & \multicolumn{1}{c}{---} & \multicolumn{2}{S[table-format=1.2(2)]}{1.50 \pm 0.04} \\
   \bottomrule
  \end{tabularx}
\end{table}

Table~\ref{table:alpha_meas} gathers the values of the fitting
parameter $\alpha$ measured by the NSI and XFaster pipelines at 95 and
\SI{150}{\giga\hertz} for two independent dust templates that are derived from
the \Planck 217 and \SI{353}{\giga\hertz} maps (Equation~\ref{eq:dust_temp}).\footnote{
We use effective band centers for the relevant \planck maps of 101.3, 220.6, and
\SI{359.7}{\giga\hertz}. These are computed from the spectral response functions of \planck\rq s polarization sensitive bolometers for a flat-spectrum source~\citep{planck2013_spectralresponse}. }  For NSI, the error in $\alpha$
for the 353-\SI{100}{\giga\hertz} template is dominated by the contribution from
chance correlations between the dust and CMB $EE$ components.  Because
XFaster includes both $EE$ and $BB$ spectra in the $\alpha$ fits, this chance correlation is subdominant; instead,
the measurement of $\alpha_{95}$ (the value of $\alpha$ appropriate to cleaning a \SI{95}{\giga\hertz} map) 
is limited by \spider noise, while the error on $\alpha_{150}$
is contributed in equal parts by \spider noise and \Planck
noise in the template.  For the 217-\SI{100}{\giga\hertz} template, the template
noise is a larger contributor to the $\alpha$ error at both
frequencies; it nearly equals the \spider noise contribution at \SI{95}{\giga\hertz}
and is 3 times more significant than \spider noise at \SI{150}{\giga\hertz}.
The use of both $EE$ and $BB$ spectra over the full multipole range
accounts for XFaster's significantly smaller uncertainty in the
determination of $\alpha$ (by a factor of 2 for
$\nu_0=\SI{353}{\giga\hertz}$, and 3.5\,--\,4 for
$\nu_0=\SI{217}{\giga\hertz}$) compared to NSI.

As can be inferred from Table~\ref{table:alpha_meas}, we detect polarized dust
emission at high significance at both 95 and \SI{150}{\giga\hertz}.
The NSI and XFaster pipelines provide consistent $\alpha$ values,
which are also broadly consistent with \Planck expectations. Assuming
a modified-blackbody dust SED with temperature
$T_\mathrm{d}=\SI{19.6}{\kelvin}$, \citet{Planck_2018_XI} find the polarized
emission from interstellar dust over \SI{71}{\percent} of the
high-Galactic-latitude sky to be consistent with a dust spectral index
$\beta_\mathrm{d}=1.53\pm 0.02$, corresponding to the $\alpha$ values
reported in the \Planck columns of Table~\ref{table:alpha_meas}.
All \spider values are within $2\sigma$ of those estimated by \planck.

\subsection{SMICA}
\label{sec:smica}

SMICA \citep[Spectral Matching Independent Component Analysis;][]{Delabrouille2003,SMICA} is a harmonic space
component separation technique.
In brief, the approach involves the calculation of the cross-spectra that preserve the joint correlation structure between the input maps.  The power in these spectra is then partitioned among individual components based on their spectral shape.
The fitted spectral components uniquely determine the weight assigned to each map and allow recovery of component-separated maps.
The formalism behind SMICA has been discussed in other publications; below we summarize the method as implemented for \spider.

The SMICA pipeline is highly complementary to the template methods above.
As implemented in this work, SMICA adopts a rigid model for
the spectral energy density of the dust foreground, but, unlike the template methods, assumes relatively little
about its spatial morphology.
This modeling flexibility comes at the cost of a larger number of fit parameters for a given set of input data.
The SMICA pipeline also enables a fully consistent joint analysis of the \planck and \spider data.  

\subsubsection{Implementation}

\spider's implementation of SMICA takes as input $N_{\text{chan}}$
polarized maps and uses PolSpice to compute the spectral covariance
matrix $\widehat{\pmb{R}}_b$: a $2N_{\text{chan}} \times
2N_{\text{chan}}$ matrix for each bandpower $b$, gathering all possible combinations of binned
$EE$ and $BB$ auto- and cross- pseudo-spectra. We then construct a parameterized model covariance $\widetilde{\pmb{R}}_b$, also in pseudo-spectrum space, that accurately describes the data $\widehat{\pmb{R}}_b$.

While not required by the approach, in this work we assume that the dust polarization amplitude follows a modified-blackbody frequency scaling whose index is scale-independent and identical in $EE$ and $BB$.
The model for polarization spectrum $X$ ($X \in \{EE,BB\}$) is as follows:

\begin{equation}
\label{eq:smica_rmodel}
\widetilde{\pmb{R}}_b^X(\theta) =\widetilde{\pmb{N}}_b^X +  \sum_{b'} \pmb{J}_{b,b'} \left[ f^X_{b'}(\beta_d) P_{b'}^X f^X_{b'}(\beta_d)^T + C_{b'}^X \right]\,,
\end{equation}
where $f^X_b(\beta_d)$ is a vector of size $N_{\text{chan}}$ that captures the dust amplitude scaling in the map domain, $P_b^X$ the full-sky bandpowers of the dust at a fixed reference frequency, $C_b^X$ the full-sky CMB bandpowers, and $\widetilde{\pmb{N}}_b^X$ is a $N_{\text{chan}} \times N_{\text{chan}}$ matrix representing the auto-correlated noise terms for all inputs.
These are all free parameters fit in the model, notated together by $\theta$ for brevity.
The transfer matrix, as introduced in Section~\ref{subsection:NSI}, is represented by $\pmb{J}_{b,b'}$ and is applied only to the terms that contribute to the sky signal.
This matrix includes effects from filtering, beam smoothing, and the mode-coupling kernel from the mask.

The model is fit by finding optimal parameters $\theta$ that minimize the spectral mismatch between the data $\pmb{\widehat{R}}_b$ and the model $\widetilde{\pmb{R}}_b(\theta)$.
This optimization fits all parameters simultaneously to account for covariance between bins and different components.
The test statistic is the Kullback-Leibler divergence between data and model:

\begin{align}
\label{eq:smica_kl_divergence}
-2\ln L &=
\sum_{b} w_b\,
\mathrm{Tr}\left[\widehat{\pmb{R}}_b \widetilde{\pmb{R}}^{-1}_b(\theta) -
  \ln \left(\widehat{\pmb{R}}_b \widetilde{\pmb{R}}^{-1}_b(\theta)\right)\right]\,, \\
w_b &= \sum_{\ell \in b} (2\ell + 1) f_{sky}\,.
\end{align}
The component-separated bandpowers are recovered by maximizing Equation~\ref{eq:smica_kl_divergence} using the Markov Chain Monte Carlo (MCMC) solver \texttt{emcee}~\citep{emcee}.

This particular choice of model parameterization and likelihood presents some subtleties in implementation.
The likelihood presented in Equation~\ref{eq:smica_kl_divergence} doesn't account for increased uncertainty of the bandpower estimates due to time-domain filtering. This factor is difficult to compute analytically, so this correction is instead applied to the chains after the fact. The covariance $\pmb{\Sigma}'$ of the component-separated bandpowers is modified to increase the uncertainty of the CMB signal bandpowers: $\pmb{\Sigma} = \pmb{B} \pmb{\Sigma'} \pmb{B}^T$. Here $\pmb{B}$ is a diagonal matrix that is determined through simulations and is insensitive to the input sky model.
Both the cleaned spectra points in Figure~\ref{fig:clean_spectra} and the SMICA likelihood in Section~\ref{sec:smica_like} include this correction factor. 

An additional challenge arises from noise fitting.
SMICA performs component separation based upon spectral shape.
Unfortunately, the dust component and noise have similar scalings, each increasing with frequency.
This introduces a degeneracy within the model (Equation~\ref{eq:smica_rmodel}) between the dust index $\beta_d$ and the noise parameters.
In order to break this degeneracy, the noise is pinned at the highest-frequency map (generally \Planck \SI{353}{\giga\hertz}) using a noise model.
This can be constructed in one of two ways: by taking the ensemble average of FFP10 noise simulations\footnote{
    \planck end-to-end ``full focal plane'' simulations \citep{planck2018_mission}},
or from the difference between the auto-spectrum of the full-mission map and the cross-spectrum of two half-mission maps.
Both noise options are propagated through the full analysis.
CMB components are largely unaffected by any particular choice of a noise model, as their spectral shape is very different from either foregrounds or noise.

The SMICA inputs are the \Spider maps in four chunk sets for each of the two frequencies (95 and \SI{150}{\giga\hertz}), and full mission \Planck HFI polarized maps (100, 143, 217, and \SI{353}{\giga\hertz}).
\WMAP and LFI channels can be incorporated to add sensitivity at lower frequencies, but were omitted from this analysis; since the \spider data are consistent with being dominated by dust foregrounds (Section~\ref{sec:a_la_Choi}), excluding these data allows for a simplified fit to a single-component foreground model.
The \Planck maps are masked to the \spider observation region, re-observed, and smoothed by the \spider 150 beams.
This preprocessing ensures that foreground modes that are filtered out in the \Spider low-level pipeline are also filtered out in the \Planck maps, which is important when constructing signal or dust component-separated maps.

\subsubsection{SMICA Results}
\label{subsubsection:SMICAfgresults}

SMICA-derived fits for the dust spectral index are shown in Table~\ref{table:alpha_meas}.
As expected, recovery of the dust spectral index is somewhat sensitive to the choice of the noise model:
indices derived from the two choices of \SI{353}{\giga\hertz} noise differ at a mutual $\sim$1$\sigma$ level.
While important to SMICA's ability to determine dust properties, the practical effect of this on the CMB component is mitigated by the correlation within the fit between $\beta_d$ and the dust amplitude $P_b$.

Because of this correlation, the dust power propagated to lower frequencies is relatively insensitive to the choice of the noise model: the dust bandpowers at \spider frequencies agree to within a mutual $\sim$0.25$\sigma$ between the two models.

The SMICA-fitted $\beta_d$ is also generally consistent with that from the template methods. 
NSI has larger $\alpha$ error, which allows for good agreement with either of SMICA's noise configurations.
XFaster's 353-\SI{100}{\giga\hertz} template value shows good agreement with the SMICA auto--cross noise model (within a mutual 0.4$\sigma$), but the consistency decreases with the FFP10 noise model (mutual 1.5$\sigma$).
While we have no \emph{a priori} reason to favor one SMICA noise model over the other,
for reasons unrelated to the foreground estimate (discussed in Section~\ref{sec:smica_like}) auto--cross was chosen as the baseline noise configuration.

\subsection{Cleaned Power Spectra}
\label{sec:clean_spectra}

\begin{figure*}
  \centering
  \includegraphics[width=\textwidth]{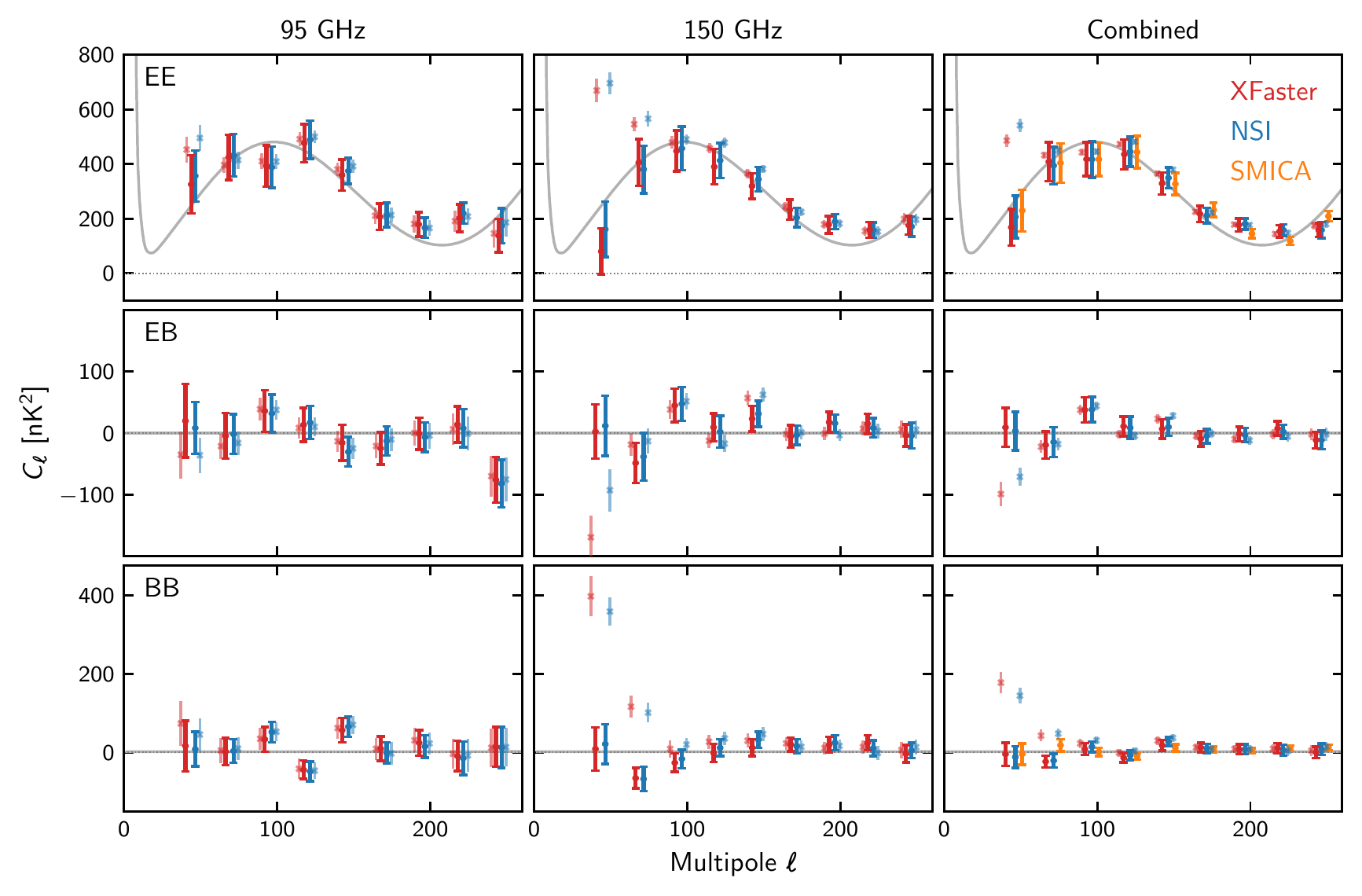}
  \caption{Foreground cleaned power spectra from the XFaster, NSI, and SMICA
    pipelines.  Foreground templates for XFaster and NSI are constructed as
    described in Section~\ref{sec:template_sub} assuming $\nu_0 = \SI{353}{\giga\hertz}$, scaled by best-fit factors
    listed in Table~\ref{table:alpha_meas}, and subtracted from the \spider\
    maps prior to computing spectra with the two pipelines.  The raw spectra
    from Figure~\ref{fig:raw_spectra} are also shown ($\times$), along with the
    best fit \planck\ $\Lambda$CDM power spectrum from \cite{planck15_params}
    (\textit{gray line}). The SMICA points are the best fit component separated
    CMB spectrum as described in Section~\ref{sec:smica}. Error bars
    for the foreground-cleaned data include sample variance for all pipelines.
    }
   \label{fig:clean_spectra}
\end{figure*}

Power spectra cleaned with the map-based (XFaster and NSI) and harmonic-space (SMICA) methods are presented
in Figure~\ref{fig:clean_spectra}, alongside the raw spectra from Figure~\ref{fig:raw_spectra} for comparison. The foreground-cleaned data points from all three pipelines include sample variance and the noise estimated by each pipeline.

All pipelines remove significant foreground power in the low $\ell$ bins, yielding $EE$ spectra in good agreement with one another and with the $\Lambda$CDM model. The biggest deviation occurs for the $\ell \sim 200$ bin, where the map-based approach fluctuates high.
In SMICA this particular bin remains high, but at a lower significance due to inclusion of \planck data. When omitting \planck 100 and \SI{143}{\giga\hertz} data, this bandpower drifts up, indicating a larger power contribution from \spider than \planck.

A comparison of the cleaned $BB$ spectra to a lensed $\Lambda$CDM model derived from the best-fit
\planck\ parameters~\citep{planck15_params} yields
$\chi^2_{lensed}=9.2~(9.3)$ for the XFaster (NSI) bandpowers.  This is an improvement of
$\Delta\chi^2=0.7~(1.4)$ over a model without lensing.
The SMICA CMB $BB$ spectrum also prefers lensing, with $\chi^2_{lensed}=9.2$ and an improvement of $\Delta\chi^2=4.0$ over an unlensed model.
The biggest difference between the template-based pipelines and SMICA occurs in the second multipole bin of the $BB$ spectrum, where the template methods fluctuate low while SMICA fluctuates high. The origin and impact of this difference on the cosmological results are discussed in Section~\ref{sec:likelihoods}.

\subsection{Polarized Synchrotron Emission}
\label{sec:a_la_Choi}

Both the template-subtraction method
(Section~\ref{sec:template_sub}) and SMICA (Section~\ref{sec:smica})
assume that the sky signal contains only one polarized foreground
component: interstellar dust.
Polarized synchrotron emission is known to be significant, even
at high Galactic latitudes, at the lower frequencies mapped by CMB
experiments \citep[see, \emph{e.g.},][for an early
measurement]{Page_etal_2007}. For \spider's region and frequencies of interest, 
however, this emission is expected to be
subdominant to that from Galactic dust. In this section, we present
the results of a harmonic-domain foreground-separation analysis that
assumes the presence of polarized Galactic dust \emph{and} synchrotron
emission in the \spider data and constrains their relative power.

Following \citet{Choi+Page}, we construct the ensemble of $EE$ and
$BB$ power spectra made up of all possible cross- and auto-spectra, computed with PolSpice,
between \wmap, \Planck HFI, and \spider maps. To avoid noise bias, each ``auto''-spectrum
is computed as the cross-spectrum between two maps constructed from years 1-5/6-9 (\wmap), half-missions (\Planck),
and interleaved sets of 10-minute data chunks (\spider; see
Section~\ref{sec:pse}). Before spectrum estimation, all \Planck
and \wmap maps are re-observed and all maps are corrected for T-to-P leakage and smoothed to 
a common $1^{\circ}$ resolution, corresponding to the resolution of the 
\wmap K-band map. When correcting power spectra, we use the unbinned filter transfer function, as the effects other than filter attenuation are highly subdominant to other sources of error.

Under the assumption that the foreground
signal in each map is made up of Galactic dust and synchrotron
emission, each spectrum in the ensemble is the sum of three physical
components (CMB, dust, and synchrotron) and their correlations. We
model the spectral energy distributions
(SEDs) of each at a given frequency $\nu$ as follows:
\begin{equation}
  \label{eq:sed_comps}
  \begin{aligned}
    I_\mathrm{CMB} & = A_\mathrm{CMB}\frac{\delta B_{\nu}}{\delta T}\Bigr|_{T_\mathrm{CMB}} \,, \\
    I_\mathrm{s}(\nu) & = A_\mathrm{s}\left( \frac{\nu}{\SI{23}{\giga\hertz}} \right)^{\beta_\mathrm{s}}\,, \\
    I_\mathrm{d}(\nu) & = A_\mathrm{d}\left( \frac{\nu}{\SI{353}{\giga\hertz}} \right)^{\beta_\mathrm{d}}
      \frac{B_\nu(T_\mathrm{d})}{B_{\SI{353}{\giga\hertz}}(T_\mathrm{d})}\,.
  \end{aligned}
\end{equation}
$A_\mathrm{CMB}$, $A_\mathrm{s}$, and $A_\mathrm{d}$ are the
amplitudes of the three components 
(in \si{\micro\kelvin\of{CMB}} for $A_\mathrm{CMB}$ and MJy for $A_\mathrm{s}$ and $A_\mathrm{d}$) referenced at 23 and \SI{353}{\giga\hertz} for synchrotron and dust, respectively.
$\beta_\mathrm{s}$ and $\beta_\mathrm{d}$ are the synchrotron and dust
spectral indices, $B_\nu(T_\mathrm{d})$ is the Planck function
computed at the dust temperature $T_\mathrm{d}=\SI{19.6}{\kelvin}$, and $\frac{\delta B_{\nu}}{\delta T}\Bigr|_{T_\mathrm{CMB}}$ is its derivative with respect to $T$ computed at $T_\mathrm{CMB} = \SI{2.7}{\kelvin}$.
In this formalism the contribution of a given component to the
cross-spectrum between two frequencies $\nu_1$ and $\nu_2$ 
is simply the product
$I_\mathrm{c}(\nu_1)\,I_\mathrm{c}(\nu_2)\,\Sigma_\mathrm{c}$, where
the index $\mathrm{c}$ labels the component and $\Sigma_\mathrm{c}$ is
the cross-spectrum of the associated $(Q, U)$ component spatial templates 
multiplying the SEDs. Similarly, a
correlation between two components $\mathrm{c}_1$ and $\mathrm{c}_2$
yields a contribution $I_{\mathrm{c}_1}(\nu_1)\,I_{\mathrm{c}_2}(\nu_2)\,\Sigma_{\mathrm{c}_1\times \mathrm{c}_2}$, where the cross-spectrum
$\Sigma_{\mathrm{c}_1\times \mathrm{c}_2}$ can be interpreted as a
scale-dependent spatial correlation coefficient between the two
components.

We perform an MCMC analysis to fit, independently in each multipole bin
and for each polarization ($EE$ or $BB$), a model to this cross-spectrum ensemble 
consisting of the spectral indices
$\beta_\mathrm{s}$ and $\beta_\mathrm{d}$, the three parameters
$\widetilde{A}_\mathrm{c}\equiv A_\mathrm{c}\sqrt{\Sigma_\mathrm{c}}$, and
the correlation coefficients
$\delta\equiv \Sigma_{\mathrm{d}\times \mathrm{CMB}}$ and
$\rho\equiv \Sigma_{\mathrm{d}\times \mathrm{s}}$.
We use broad, uniform priors on nearly all components, excepting $\beta_\mathrm{s}$, where we use \Planck's posterior of $\beta_\mathrm{s} = -1.15 \pm 0.17$ \citep{Planck_2018_XI}, and $\delta$, where we use the distribution of correlations observed between a large number of CMB- and dust-only sky simulations. There is no \emph{a priori} reason to expect more than chance correlation between the dust and CMB components, and so it is reasonable to measure $\delta$ in this fashion. Unlike the CMB, the correlation between the dust and synchrotron emission (parameterized by $\rho$) is expected to be non-zero.  We adopt a uniform prior on this parameter.
Note that we do not fit for the
$\Sigma_{\mathrm{s}\times \mathrm{CMB}}$ correlation coefficient,
expected to be the smallest of the three, as the validation of this
method on simulations did not indicate a need for it.
Errors on the input spectra are computed from a distribution of signal and noise simulations, with each map having a noise model appropriate for its associated instrument: 
 \Spider's stationary noise model (without the noise model scaling factor), \Planck FFP10 simulations, and Gaussian pixel noise for \WMAP.

\begin{figure}
  \centering
  \includegraphics[width=\columnwidth]{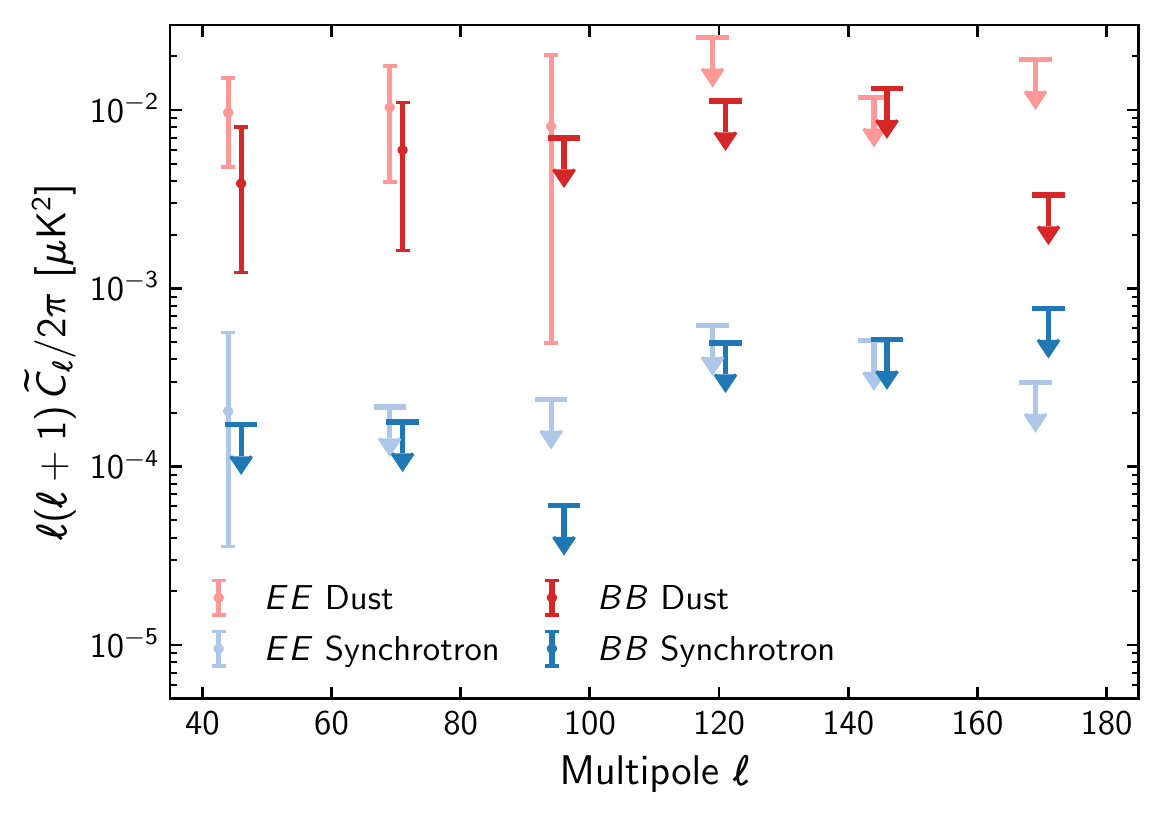}
  \caption{\SI{95}{\giga\hertz} $EE$ (\textit{light}) and $BB$ (\textit{dark}) angular power spectra
  of the synchrotron (\textit{blue}) and dust (\textit{red}) emission in the \spider
  region as estimated by the foreground separation method described in
  Section~\ref{sec:a_la_Choi}. These power spectra are \emph{not}
  corrected for the \spider beam or filtering. In each multipole bin,
  bandpower points are reported as the maximum likelihood point of the bandpower posterior, with the narrowest \SI{95}{\percent} of samples about the maximum likelihood defining error bars. For bandpowers which contain 0 in the narrowest \SI{95}{\percent} of points, we report \SI{95}{\percent} upper~limits.
  }
  \label{fig:sed_ratios}
\end{figure}

Drawing from the posteriors of the fitted parameters, we compute the
$EE$ and $BB$ power spectra of the dust and synchrotron emission
at \spider frequencies. The \SI{95}{\giga\hertz} bandpowers, shown in
Figure~\ref{fig:sed_ratios}, are shown as the maximum likelihood point of the posteriors, with error bars bounding the narrowest \SI{95}{\percent} of the posterior about the maximum likelihood. For those bins in which 0 is within that \SI{95}{\percent}, only the upper limit is shown. In each multipole bin with a reported maximum likelihood for dust, the \SI{95}{\percent} upper
limit on the polarized synchrotron emission is an order of
magnitude or more below the maximum likelihood of the dust bandpower
distribution. The foreground analyses presented in
Sections~\ref{sec:template_sub} and~\ref{sec:smica}, which measure
dust levels that are comparable to those shown in
Figure~\ref{fig:sed_ratios}, therefore assume the presence of only one
polarized Galactic emission in the \spider and \Planck HFI maps, that
from interstellar dust.  Note that the power reported in
Figure~\ref{fig:sed_ratios} is \emph{not} corrected for the \spider beam
or filter transfer function; the on-sky foreground power will be
discussed in a forthcoming publication.

\subsection{Discussion}
The map-based (template-fitting) and harmonic-domain (SMICA) component separation techniques agree well in their estimation of dust model parameters and CMB spectra.
The template-fitting method imposes no model for dust's spectral scaling, and the SMICA method makes no assumptions about the spatial distribution of dust.
Thus, this agreement is evidence that their differing assumptions are valid for the \spider data set in combination with \planck.
Further, the level of synchrotron emission is constrained to be well below that of dust in this region of sky at \spider frequencies.
This justifies the assumption of dust-dominant foregrounds made in the template-fitting and SMICA pipelines.
Having established confidence in and consistency among the methods for calculating CMB component power spectra, we then use each to construct likelihoods for the tensor-to-scalar ratio, $r$.

\section{Likelihoods}
\label{sec:likelihoods}
In this section we use the power spectrum estimates above to 
construct likelihoods for the tensor-to-scalar ratio, $r$.
Two separate approaches are taken to construct and sample from the parameter likelihood.
The XFaster likelihood construction is Gaussian in the $a_{\ell m}$
coefficients, and the algorithm can naturally be adapted to sample that likelihood as
a function of parameters other than bandpowers.
This approach relies solely on the assumption that the cleaned
maps are dominated by CMB and noise, both well-approximated as Gaussian random fields.
The NSI and SMICA methods proceed instead from the bandpowers computed above, 
assuming Gaussian likelihoods for the CMB bandpowers in order to sample additional parameters.
This approach is susceptible to sample
variance, which limits the validity of the Gaussian approximation in the presence of significant $E$-mode or $B$-mode power \citep{bond2000}.

The two approaches also differ significantly in the way that parameter covariance is propagated.
The XFaster approach samples a likelihood of the $a_{\ell m}$s as a
function of the three-dimensional parameter space of $r$ and $\alpha$;
this is then marginalized into a final posterior for $r$.
The NSI and SMICA approach samples a profile likelihood in $r$,
which optimizes over foreground parameter dependence in a separate step.
All other $\Lambda$CDM parameters (notably $A_s$ and $\tau$) are held fixed.

Each of the methods described in
this work has been extensively validated on simulations that include
the cosmological signal, a model of Galactic foregrounds, and
time-domain instrumental effects, including the noise and in-flight
pointing.
Subject to the assumptions made regarding these inputs, we
find all estimators to be free from bias.
We adopt the XFaster pipeline as our baseline, given its generality and self-consistency.

\subsection{XFaster}
\label{sec:like_XFaster}

The XFaster power spectrum estimator can naturally provide a likelihood for a parameterized model directly, rather than fitting for the maximum likelihood bandpower deviations.
The generalized XFaster parameter likelihood has a form similar to that in Equation~\ref{xf_cl_likelihood}:
\begin{equation}
  -2\ln L(\theta|\widetilde{\pmb{d}}) = \sum_{\ell, k} (2 \ell + 1) g_\ell^k \left[\widetilde{\pmb{C}}_\ell^{-1}(\theta)\,\widehat{\pmb{C}}^{\,}_\ell + \ln \,\widetilde{\pmb{C}}^{\,}_\ell(\theta)\right]_{kk},
\end{equation}
where $\theta$ is a set of parameters, $\widetilde{\pmb{C}}_\ell$ is the model pseudo-$C_\ell$ matrix,
$\widehat{\pmb{C}}_\ell$ is the data pseudo-$C_\ell$ matrix, $g_\ell$ is the
mode count recalibration factor, and the index $k$ labels the maps
used (for this paper, four \SI{95}{\giga\hertz} and four \SI{150}{\giga\hertz} maps).
The likelihood is based on the XFaster approximation of the likelihood
for the observed pseudo-$a_{\ell m}$s and is therefore also Gaussian
without loss of generality.
The model pseudo-$C_\ell$s are computed using the same transfer functions, beam window functions, and mode-coupling kernels used for the bandpower computation, with the tensor contributions to the $EE$ and $BB$  power spectra modeled as a function of $r$.

Foreground fitting could, in principle, be accomplished by adding a scaled dust template spectrum to the model and fitting for the scale factor $\alpha$ at each frequency.
However, because all terms in the signal model are treated as Gaussian random fields, XFaster's estimated error would include sample variance proportional to the foreground amplitudes, which is not appropriate for the fit to a non-Gaussian template.
Therefore we instead subtract the scaled template from the data at the map level and model the residuals as CMB and noise.
Because the template subtraction is not accounted for in the covariance, the additional error from foreground cleaning must be calibrated using an ensemble of CMB, noise, and template simulations.

The data terms after template subtraction are
\begin{equation}
  \widehat{\pmb{C}}^{i\times j} = \left<(\pmb{d}_i - \alpha_i \pmb{t}_i) \times (\pmb{d}_j - \alpha_j \,\pmb{t}_j)\right>,
\end{equation}
where $\pmb{d}$ is a data map, $\pmb{t}$ is a template map, $\alpha$ is the frequency-dependent template scaling, and $i$ and $j$ are indices of the eight maps.
The template maps used for each cross-spectrum are separate \planck
half-missions to avoid contributions from \planck noise auto-spectra.
In order to compute the maximum likelihood parameter estimates, an MCMC sampler steps through values of $r$, $\alpha_{95}$, and $\alpha_{150}$, recomputing the data and model terms in the likelihood at each step.

Monte Carlo simulations of this process are used to propagate the uncertainty in
template fitting to the likelihood derived from the data.  The same process is repeated for 300 simulations using different realizations of CMB, \spider noise, and \planck noise from the FFP10 simulation ensemble.
The parameter distributions recovered from these simulations are used to estimate the added covariance from uncertainty in the template fit.
The magnitude of the additional covariance was found not to change significantly within reasonable ranges of $r$ or $\alpha$, or with different morphologies of the simulated foregrounds.
We incorporate this added uncertainty to the data's parameter likelihoods by adding it to each Monte Carlo sample.
The terms contributing to the additional covariance are noise in the template (\SI{45}{\percent}), chance correlations between \spider noise and the template (\SI{45}{\percent}), and chance correlations between the CMB and the template (\SI{10}{\percent}), where the total quadrature 1$\sigma$ error added for $r$ is 0.10.
This accounts for approximately half of the total error.
Uncertainties in the beam window functions and in the corrections to the \spider noise model are also parameterized and marginalized over in the final result; their effects are negligible.

\begin{figure}
  \centering
  \includegraphics[width=\columnwidth]{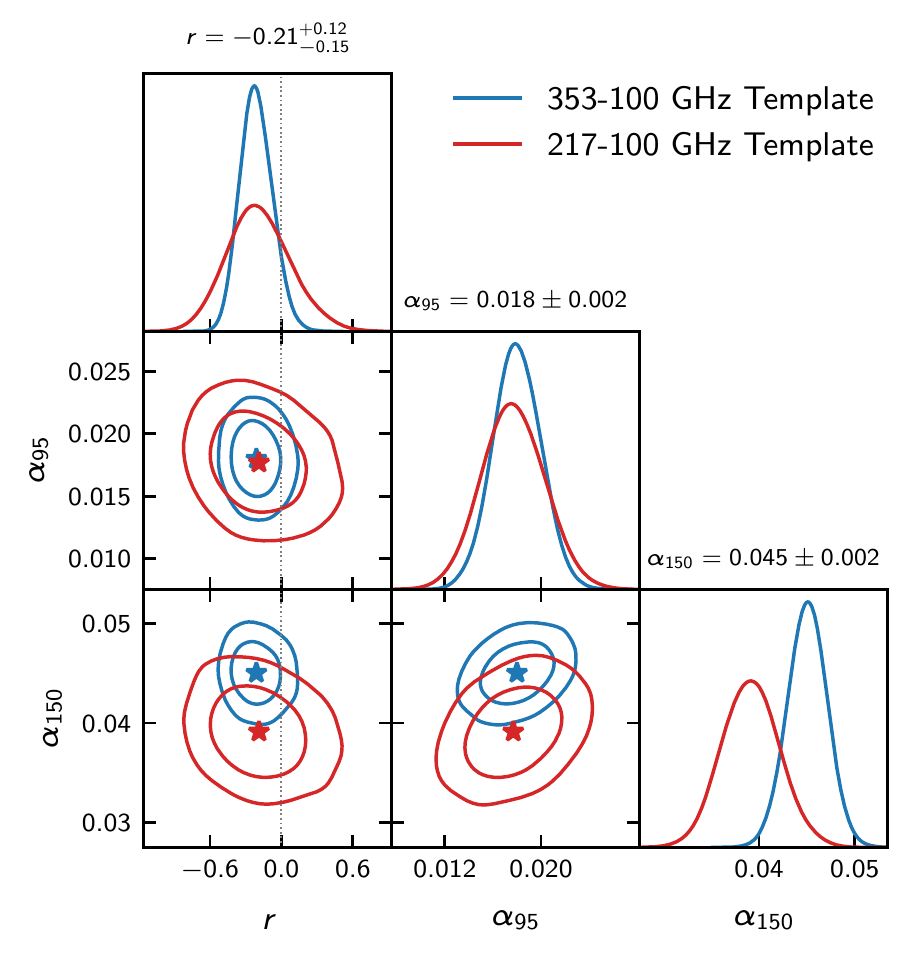}
  \caption{The combined XFaster likelihood for $r$ and $\alpha$, imposing no priors on these parameters.
    Blue curves show likelihoods computed using a 353-\SI{100}{\giga\hertz} template.
    Red shows the same for a 217-\SI{100}{\giga\hertz} template, where $\alpha$ values have been scaled assuming a modified blackbody for dust ($T_d=\SI{19.6}{\kelvin}$, $\beta_d$ determined from each $\alpha$ sample) to the corresponding values for a 353-\SI{100}{\giga\hertz} template.
    1$\sigma$ constraints for the 353-\SI{100}{\giga\hertz} template are shown in the panel titles.}
  \label{fig:like_xf}
\end{figure}

Figure~\ref{fig:like_xf} shows the complete XFaster likelihood result using both choices of dust template (353-\SI{100}{\giga\hertz} and 217-\SI{100}{\giga\hertz}), incorporating error contributions from the template subtraction and with no priors imposed on these parameters.
The scaling of the 217-\SI{100}{\giga\hertz} template $\alpha$s to plot on common axes relies on the assumption of a modified-blackbody dust model.
The two templates yield consistent results for all parameters.
Due to its greater constraining power, we use the 353-\SI{100}{\giga\hertz} template result for our final constraint, yielding a maximum-likelihood estimate for the tensor-to-scalar ratio of $r_{mle}=-0.21$.
We find that \SI{6}{\percent} of simulations with input $r=0$ yield $r_{mle}<-0.21$, so such a value is not inconsistent with expected noise fluctuations.

We can compute an upper limit on $r$ from this likelihood, subject to the physical constraint that $r\geq 0$.
\textcolor{black}{Imposing a flat prior on $r$, truncated for $r<0$ to implement this physical constraint,
we obtain a \SI{95}{\percent} Bayesian upper limit of $r<0.19$.}

\begin{figure}
  \centering
  \includegraphics[width=\columnwidth]{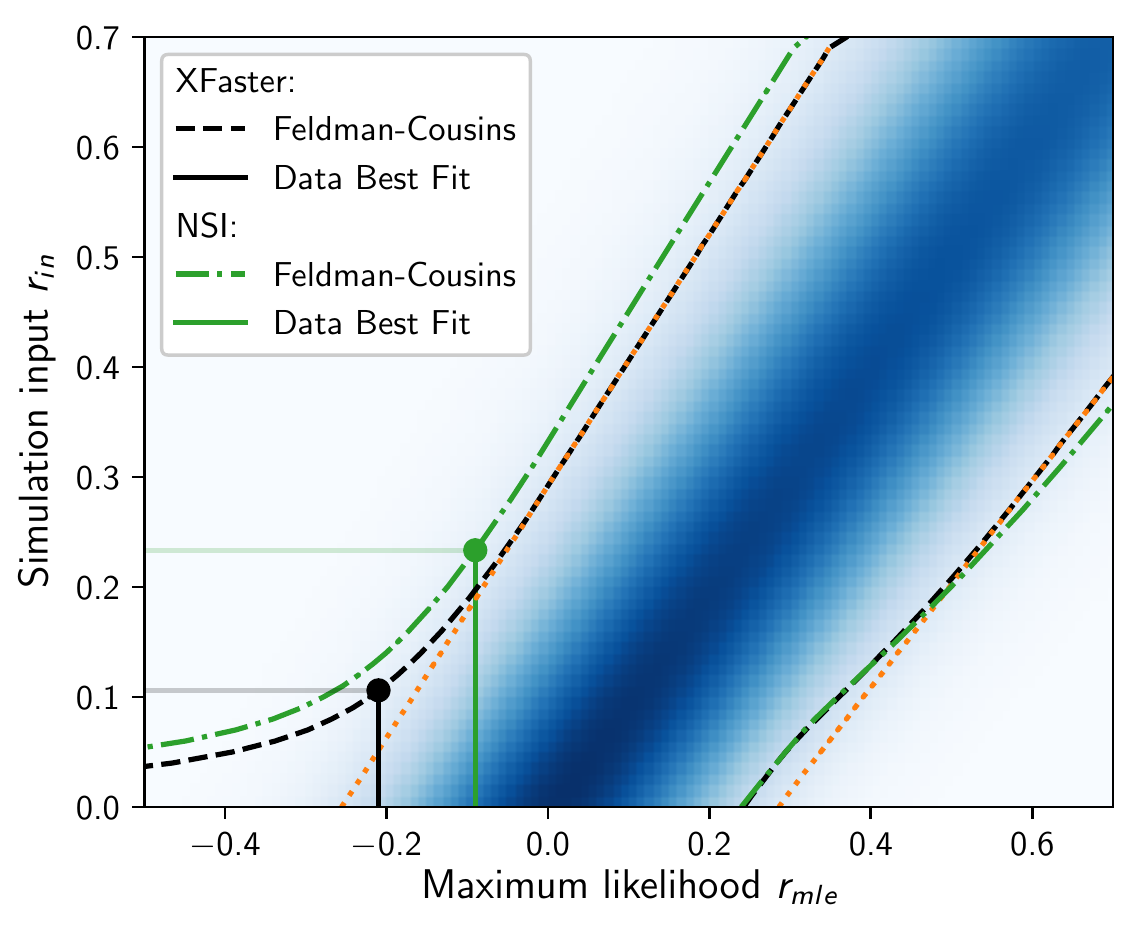}
  \caption{Feldman--Cousins \SI{95}{\percent} confidence interval (\textit{dashed black}) 
     on $r$ as a function of $r_{mle}$,
     derived from template-subtracted XFaster likelihoods.
     The observed $r_{mle}$ is indicated by the black vertical line;
     our upper limits are the intersections of the curves above with this line,
     projected onto the vertical axis.
     Blue shading indicates the distribution of $r_{mle}$ as a function of input $r$, which is used to construct these curves.
     For each input $r$ we conduct 300 XFaster simulations to produce a histogram of $r_{mle}$,
     which we smooth by fitting a Gaussian model (a good fit).
     At each input $r$, \SI{95}{\percent} of simulation results lie between the dotted orange lines.
    These simulations use noise maps rescaled in each bandpower bin by the associated
    XFaster estimate of the noise model scaling factor.
    Similarly, the Feldman--Cousins interval (\textit{green dash-dot}) and data best fit
    (\textit{green solid}) are shown for the NSI pipeline; the associated simulations are not shown.
  }
  \label{fig:fc}
\end{figure}

We also construct a classical confidence interval for $r$, following the approach
discussed in~\cite{Feldman_1998}.
\textcolor{black}{In this approach, simulations are conducted for a range of values of input $r_{in}$, each carried through to a value of $r_{mle}$.
For each $r_{in}$ an interval of $r_{mle}$ is defined containing \SI{95}{\percent} of simulations---those with the largest values of the likelihood ratio 
$R\equiv\mathcal{L}(r_{mle}|r_{in})/\mathcal{L}(r_{mle}|r^*)$, where $r^*$ is the value of $r_{in}$ that maximizes 
$\mathcal{L}(r_{mle}|r_{in})$. Note that $r^*=0$ for $r_{mle}\leq0$.
Figure~\ref{fig:fc} shows this confidence interval as black dashed lines,
which transition smoothly between detection and upper limit while
maintaining correct coverage for $r_{mle}$ near or beyond the physical boundary $r \ge 0$.
The observed $r_{mle}$ yields an upper limit of $r<0.11$ (\SI{95}{\percent} CL).}
\textcolor{black}{The difference between this and the Bayesian limit reflects their disparate definitions and interpretations,
as well as the modest over-coverage (conservatism) of the Bayesian limit near the physical boundary.}

An ensemble of CMB, \spider noise, and template noise simulations are used to
determine relative contributions to the error budget. Simulated maps are
constructed by creating an ensemble in which only one of these
components is allowed to vary (\emph{e.g.}, 300 maps made by joining a single
CMB realization, a single \spider noise realization, and 300 template
realizations).  By comparing the scatter in the estimated $r$ values,
we can estimate the relative contributions to the total error, \emph{i.e.}, the
scatter when all three components are varied together.  Assuming the template is a
perfect representation of the dust morphology, the largest contributor to $\sigma_r$ is
\spider noise, including its chance correlations with the template, at
$\sim$\SI{60}{\percent}. CMB sample variance and chance correlations contribute
$\sim$\SI{25}{\percent}, and template noise adds $\sim$\SI{15}{\percent}.
For the 217-\SI{100}{\giga\hertz} template,
the statistical error is instead dominated by the noise in the
template.

\subsection{NSI}

The foreground-cleaned power spectra from the NSI pipeline are also propagated to an $r$-likelihood.
This likelihood proceeds in separate steps: the bandpowers are first estimated, then fit for the foreground template amplitudes $\alpha$  (as in Section~\ref{sec:template_sub}), and finally an $r$-likelihood is constructed from the cleaned spectra.
As a conservative precaution to avoid bias from foregrounds on $r$, the fit for $\alpha$ uses only $EE$ while the fit for $r$ uses only $BB$.
This fit for $r$ uses the simple Gaussian likelihood:
\begin{multline}
    -2 \ln L = \\ \left(\widehat{C}_{b}^{BB} - C_{b}^{BB}(r)\right)^T \pmb{\textrm{M}}^{-1} \left(\widehat{C}_{b}^{BB} - C_{b}^{BB}(r)\right) + \ln \vert\pmb{\textrm{M}}\vert,
\label{eq:nsi_r_like}
\end{multline}
where $\widehat{C}_{b}^{BB}$ is the cleaned $B$-mode spectrum measured by \spider and $C_{b}^{BB}(r)$ is a $\Lambda$CDM model using \planck parameters, lensing, and allowing $r$ to vary.
The bandpower covariance matrix $\pmb{\textrm{M}}$ is a sum of three contributions: \spider's statistical noise, estimated from the distribution of 378 NSI cross-spectra; sample variance, estimated from an ensemble of re-observed signal-only $\Lambda$CDM simulations; and propagated error on $\alpha$, to capture the statistical error on the foreground template fitting.
The cross-spectra between terms for signal, noise, and foreground template, while uncorrelated in the mean, also contribute to the total covariance.
These extra contributions are estimated together from an ensemble of full signal plus noise simulations.
Note that, unlike NSI results for raw power spectra and $\alpha$ fits, this result depends on simulations of \spider noise.
While bandpowers do not in general follow a Gaussian likelihood \citep{bond2000, gerbino2020}, simulations show that this approximation is adequate for \spider's sky coverage and $\ell$ bins.

The errors on $\alpha_{95}$, $\alpha_{150}$, and the associated covariance are propagated to both the bandpower covariance matrix and the $r$-likelihood using a Monte Carlo method.
Starting with the two-dimensional Gaussian distribution described by the best-fit parameters in Table~\ref{table:alpha_meas}, random $\alpha$ values are drawn, and new template-subtracted bandpowers are computed.
A set of 1000 such randomly cleaned bandpowers are used to estimate the bandpower covariance due to $\alpha$ error, which is added to $\pmb{\textrm{M}}$ as above.
To estimate $r$, Equation~\ref{eq:nsi_r_like} is evaluated 4096 times for another set of 4096 randomly cleaned bandpowers (as $\widehat{C}_{\ell}^{BB}$).
This step allows $\alpha$ to shift slightly from its best-fit value when the $r$-likelihood prefers it.
More draws are required than for the previous step (4096 vs 1000) so that the random seed does not significantly impact results.
The final $r$-likelihood is the average of the likelihoods evaluated for each random draw from the $\alpha$ distribution.

\textcolor{black}{Finally, as for XFaster, we derive upper limits on $r$ from the NSI likelihood under the physical constraint $r \geq 0$.
A Feldman--Cousins approach (Figure~\ref{fig:fc}) yields $r<0.23$, while
a Bayesian calculation yields $r < 0.27$, both at \SI{95}{\percent} confidence.}
Section~\ref{sec:like_discuss} further discusses how the NSI and XFaster results compare.

\subsection{SMICA}
\label{sec:smica_like}

As for NSI, we construct an $r$-likelihood for SMICA
under a simple Gaussian approximation for the bandpower likelihood (Equation~\ref{eq:nsi_r_like}). 
We note that this is a suboptimal approximation for the $r$-likelihood due to measured non-Gaussianity of the $BB$ bandpowers in the lowest bins.
For SMICA, the covariance $\pmb{M}$ is empirically determined from MCMC chains. 
While this covariance does not capture the full $r$-dependence of sample variance, this contribution to the total covariance is expected to be small.
Furthermore, since the SMICA likelihood maximization that produced these bandpowers jointly fits for CMB signal, instrumental noise, and dust foregrounds,
the statistical distribution of the fitted CMB bandpowers, $C_b$, naturally includes noise and foreground uncertainty.

As discussed in Section~\ref{sec:smica}, this implementation of SMICA requires a choice of noise model at \SI{353}{\giga\hertz}.
Lacking strong justification in preferring one over the other, and in the spirit of reporting a conservative upper limit, we run the analysis with both choices and report the less stringent result. Consequently, the reported SMICA results come from the auto--cross noise model.

\begin{figure}
  \centering
  \includegraphics[width = \columnwidth]{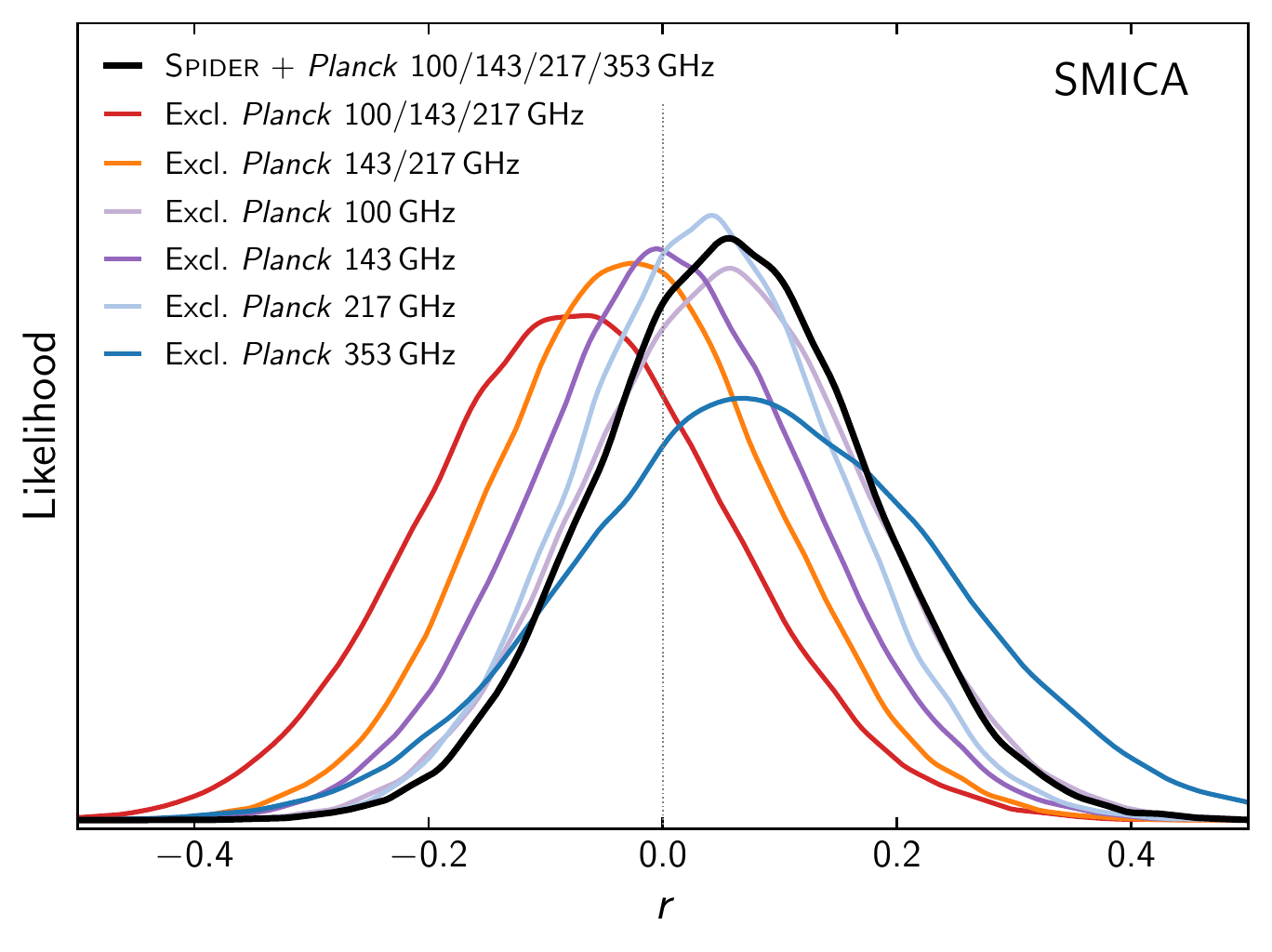}
  \caption{
    The SMICA $r$-likelihood, showing the impact of varying the inputs into the pipeline.
    Removing any of the non-353 \planck bands lowers $r_{mle}$, consistent with \spider data driving $r$ low.
    This suggests that the \spider and \planck noise, or their chance correlation with the foreground signal, must fluctuate with opposite sign.
    }
 \label{fig:smica_r_upperlimit}
\end{figure}

The resulting SMICA $r$-likelihood is shown in Figure~\ref{fig:smica_r_upperlimit}. The nominal configuration with all \spider and \planck data yields a maximum likelihood estimate of $r_{mle}=0.06\pm0.11$.
Subject to a physical prior that $r \geq 0$, this corresponds to a \SI{95}{\percent} \textcolor{black}{Bayesian} upper limit of $r < 0.24$.
\textcolor{black}{A Feldman--Cousins constraint is computationally impractical in the SMICA framework.}

In Figure~\ref{fig:smica_r_upperlimit} we further explore the effect on the SMICA likelihood of incorporating different subsets of the \planck data.
The \planck 100 and \SI{143}{\giga\hertz} data are particularly interesting in this regard, as they contribute appreciably to the SMICA CMB component but not directly to the template analysis.  At the angular scales of interest, the SMICA algorithm assigns weights to the \spider data at 95~(150)\,GHz that are approximately five (three) times those applied to the \planck data at the nearest corresponding frequency. 
Omitting the \planck data at 100 and \SI{143}{\giga\hertz} shifts the $r$-estimate downward to $r_{mle} = -0.03 \pm 0.12$, in closer agreement with the template results.
A configuration similar to the template methods (\planck 143 and \SI{217}{\giga\hertz} data omitted) results in a similar value ($r_{mle} = -0.02^{+0.12}_{-0.11}$).
Even in this configuration, however, \planck \SI{100}{\giga\hertz} still has substantial influence on CMB recovery, with a weight approximately four times that in the template methods. When omitting \planck 100, 143 and \SI{217}{\giga\hertz} data, SMICA recovers an $r_{mle}$ closest to that from the template methods: $r_{mle} = -0.07 \pm 0.13$.

In each case, we find that the shift in $r_{mle}$ is primarily driven by the first two multipole bins of the $BB$ spectrum.
This suggests that either the \spider noise or chance correlations between noise and foregrounds result in a negative fluctuation in $BB$ relative to the \planck data, irrespective of the method of foreground removal.
Variation among the $r$ estimates may also arise from the differing assumptions made regarding the modeling of the foregrounds in each pipeline.

\subsection{Discussion}
\label{sec:like_discuss}

\begin{table}
  \centering
    \caption{Summary of $r$-likelihood values from various pipelines, with nominal upper limits in bold.
    }
    \label{table:likelihood}
    \begin{tabular}{llccc}
      \toprule
      \toprule
      Pipeline & Description & $r_{mle}$ & $r \leq 95 \si{\percent} $ \\
      \midrule
      {XFaster} & {Nominal, Feldman--Cousins} & {-0.21} & \textbf{0.11} \\
      & {Nominal, Bayesian} & {-0.21} & \textbf{0.19} \\
      & NSI-like: & & \\
        & \quad ($a$) $r$ from $BB$ only & -0.19 & -- \\
        & \quad ($b$) Independent $EE$ \& $BB$ noise & -0.19 & -- \\
        & \quad ($a$) + ($b$) & -0.15 & -- \\
      \midrule
      {NSI} & {Nominal, Feldman--Cousins} & {-0.09} & {0.23} \\
      & {Nominal, Bayesian} & {-0.09} & {0.27} \\
      \midrule
      {SMICA} & {Nominal, Bayesian} & {0.06} & {0.24} \\
      & Template-like: &  &  \\
        & \quad Excl. \planck inputs $<\SI{353}{\giga\hertz}$ & -0.07 & -- \\
      \bottomrule
    \end{tabular}
\end{table}

Table~\ref{table:likelihood} presents the maximum likelihood $r$ for each of the three pipelines, both in their standard configurations and in various modified configurations chosen to explore the impact of their structural differences (discussed further below).
The nominal configuration of each pipeline was chosen prior to running the estimator on data
and includes its maximal data set---all $EE$ and $BB$ science bins for
XFaster, and the full set of \spider and \planck maps for SMICA.
The same table reports the \SI{95}{\percent} upper limit for each pipeline in its the nominal configuration.
In
all cases, XFaster, NSI, and SMICA are found to return
unbiased $r$ posteriors that
are broadly consistent with one another when run on time-domain
simulations. We adopt the XFaster pipeline as our primary result due
to the more formally correct construction of its
likelihood.

Insofar as each of the above pipelines is unbiased on simulations and makes relatively simple (and non-contradictory) assumptions, 
the observed difference in $r_{mle}$ when restricted to a closely comparable subset of data merits investigation. 
An important question is whether the observed discrepancy between methods is consistent with expected variation given the difference in methodologies alone. 
We address this question below in two ways: by observing the effects on the data's $r_{mle}$ from slight modifications of each pipeline, and (where feasible) by comparing the results of each pipeline when applied to identical simulated maps.

We first compare the two template-subtraction methods, XFaster and NSI.
XFaster's tighter upper limit results primarily from a lower $r_{mle}$, as illustrated in Figure~\ref{fig:fc}.
The NSI limit is also increased slightly by having a broader distribution than XFaster, as a result of less-optimal weighting of the available data.
Table~\ref{table:likelihood} highlights the effects of modifying some of the assumptions that differ between the two methods.
When XFaster is run in a more NSI-like configuration---fitting $r$ from $BB$ only, with independently estimated $EE$ and $BB$ noise---its nominal $r_{mle} = -0.21$ shifts to $-0.15$, in better agreement with
NSI's $r_{mle} = -0.09$. 
This shift of 0.06 in $r_{mle}$ is consistent with expectations from simulations, which show random fluctuations with $\sigma=0.06$ between the two different XFaster configurations when applied to the same simulated maps.

SMICA differs from the template subtraction methods in the quantity of \planck data that are incorporated in the analysis.
As shown in Table~\ref{table:likelihood} and Figure~\ref{fig:smica_r_upperlimit}, changing the selection of \planck data to better match the template methods results in a downward shift of $r_{mle}$ by $0.13$, in the direction of the template results, implying that fluctuations in the \spider and \planck data drive the $BB$ spectrum in opposite directions. 

\begin{figure}
  \centering
  \includegraphics[width = \columnwidth]{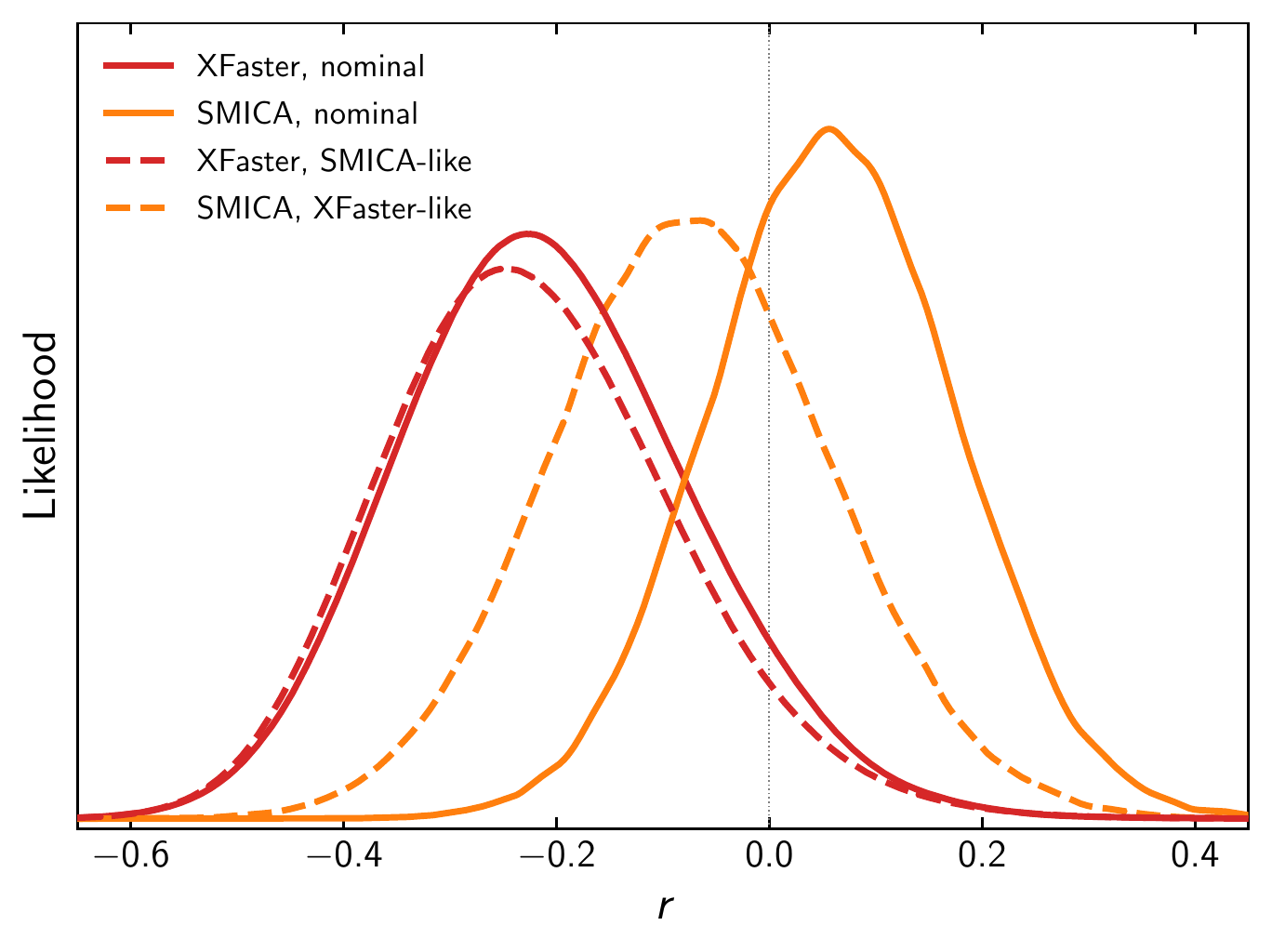}
  \caption{Comparison of the SMICA and XFaster $r$-likelihoods.
    The nominal cases correspond to the ``Nominal'' row for each in Table~\ref{table:likelihood}, using the maximal data set for each pipeline.
    Posteriors are also shown for configurations in which the pipelines use approximately the same data inputs.
    For XFaster, this corresponds to case \emph{a} in Table~\ref{table:likelihood}, limiting the $r$ fit to the $BB$ spectrum.
    For SMICA, this corresponds to the final case in Table~\ref{table:likelihood}, removing all \planck data except at \SI{353}{\giga\hertz} (reproduced from the red line in Figure~\ref{fig:smica_r_upperlimit}).
    }
  \label{fig:like_compares}
\end{figure}

The difference in $r_{mle}$ between SMICA and XFaster is not entirely resolved through the inclusion of common data products.  This is shown most clearly in Figure~\ref{fig:like_compares}. To quantify the significance of the remaining difference ($\delta r_{mle}=0.14$), we compare XFaster and SMICA in simulation by applying them to a nearly identical set of
CMB, noise, and foreground simulation maps.
For SMICA, we simulate the configuration in which the only \planck map used is \SI{353}{\giga\hertz}; for XFaster, the simulated template includes both 353 and \SI{100}{\giga\hertz} simulated \planck noise.
Both estimators recover $r_{mle}$ without bias and with partially-correlated variance. 
The covariance between the two estimators can be written as:
\begin{equation}
\pmb{\Sigma}_{\mathrm{XF,SMICA}} = \begin{pmatrix}
\sigma_\mathrm{XF}^2 & \rho \, \sigma_\mathrm{XF} \, \sigma_\mathrm{SMICA} \\
\rho \, \sigma_\mathrm{XF} \, \sigma_\mathrm{SMICA} & \sigma_\mathrm{SMICA}^2 
\end{pmatrix}. 
\label{eq:est_cov}
\end{equation}
For a 200-simulation ensemble with input $r = 0$, we find
$\sigma_\mathrm{XF} = 0.13$, $\sigma_\mathrm{SMICA} = 0.13$, and $\rho = 0.74$. 
The uncorrelated variance between the two estimators ($\rho<1$) captures the 
degree to which each is sensitive to a different
projection of the data when estimating $r_{mle}$, 
leading to statistical
variation between the methods even when given nearly identical input data;
it also suggests some degree of non-optimality in the estimators.
Comparing this ensemble to XFaster's nominal result and SMICA's template-like result in
Table~\ref{table:likelihood}, we find that the observed difference between these estimators (0.14) 
is consistent with the range of differences seen in simulations ($\sigma=0.1$).
Further, if we compare the pair of observed estimator values to the simulation ensemble in two dimensions, 
we find that about one in six of the simulated pairs result in a difference that is equal to or greater than that obtained on the data.

\section{Conclusion}
\label{sec:conclusion}

The data from \spider\rq s first flight have returned maps of the intensity and polarization at 95 and \SI{150}{\giga\hertz} that are substantially deeper than the \planck data in the same region of sky.  A rigorous suite of consistency tests have been used to define a subset of these data that can be reliably used for cosmological analysis. 
These maps, in concert with data from \planck , are used to constrain the amplitude of any cosmological $B$-mode signal in the cosmic microwave background.

As anticipated, polarized Galactic dust emission is observed with high signal-to-noise.
In \spider's sky region, the Galactic $E$-mode component has roughly twice the power of the $B$-mode component, and is found to be dominated by thermal dust emission at \SI{95}{\giga\hertz} and above at all angular scales probed;
Galactic synchrotron radiation is found to be strongly subdominant.

Separating the dust component from the cosmological signal is the
principal challenge of the present analysis.  To this end, two basic
approaches are employed: map-based template subtraction and SMICA,
an internal linear combination applied in the harmonic domain.
While the \spider and \planck $B$-mode data are found to push the constraints in opposite directions, the $r_{mle}$ derived from the template methods and SMICA are found to be consistent, subject to the assumptions made in each.
Under the assumption that our \Planck-derived template accurately captures the
morphology of the dust, \textcolor{black}{we derive the 95\% upper limit on the primordial tensor-to-scalar ratio as $r<0.11$ and $r<0.19$ using Feldman--Cousins and Bayesian approaches, respectively.}

Relaxing assumptions regarding the morphology of the dust component and
assuming a dust spectral energy distribution that is both independent of angular scale and well characterized by a modified-blackbody spectrum, SMICA gives a
\textcolor{black}{somewhat} higher upper limit of $r<0.24$.
Unlike the template-based method, this constraint is derived from a joint analysis of \spider and \planck 100-\SI{353}{\giga\hertz} data.
Further characterization of the dominant Galactic foreground emission
is the subject of a forthcoming paper.  

An improved
characterization of the foreground  emission is the focus
of \spider 's upcoming flight, which will feature a suite of three new
\SI{280}{\giga\hertz} receivers~\citep{bergman_ltd17,shaw_spie}.  These data will both complement the \planck
data at 217 and \SI{353}{\giga\hertz} and achieve significantly higher sensitivity.
At the same time, the availability of an independent data set over a
substantial portion of the full sky facilitates qualitatively new
measures of the robustness of foreground separation techniques to
choices made in the analysis and the selection of data.

\section*{Acknowledgments}
\Spider is supported in the U.S. by the National Aeronautics and Space Administration under grants NNX07AL64G, NNX12AE95G, and NNX17AC55G issued through the Science Mission Directorate and by the National Science Foundation through PLR-1043515.
Logistical support for the Antarctic deployment and operations is provided by the NSF through the U.S. Antarctic Program.
Support in Canada is provided by the Natural Sciences and Engineering Research Council and the Canadian Space Agency.
Support in Norway is provided by the Research Council of Norway.
Support in Sweden is provided by the Swedish Research Council through the Oskar Klein Centre (Contract No.\ 638-2013-8993) as well as a grant from the Swedish Research Council (dnr.\ 2019-93959) and a grant from the Swedish Space Agency (dnr.\ 139/17).
The Dunlap Institute is funded through an endowment established by the David Dunlap family and the University of Toronto.
The multiplexing readout electronics were developed with support from the Canada Foundation for Innovation and the British Columbia Knowledge Development Fund.
KF holds the Jeff \& Gail Kodosky Endowed Chair at UT Austin and is grateful for that support.
WCJ acknowledges the generous support of the David and Lucile Packard Foundation, which has been crucial to the success of the project.
CRC was supported by UKRI Consolidated Grants, ST/P000762/1,
ST/N000838/1, and ST/T000791/1.

Some of the results in this paper have been derived using the \texttt{HEALPix} package \citep{HealPix}.
The computations described in this paper were performed on four computing clusters: Hippo at the University of KwaZulu-Natal, Feynman at Princeton University, and the GPC and Niagara supercomputers at the SciNet HPC Consortium \citep{Scinet,Niagara}.
SciNet is funded by the Canada Foundation for Innovation under the auspices of Compute Canada, the Government of Ontario, Ontario Research Fund - Research Excellence, and the University of Toronto.

The collaboration is grateful to the British Antarctic Survey, particularly Sam Burrell, and to the Alfred Wegener Institute and the crew of R.V.  {\it Polarstern} for invaluable assistance with the recovery of the data and payload after the 2015 flight.
Brendan Crill and Tom Montroy made significant contributions to \Spider's development.  Paul Steinhardt provided very helpful comments regarding the status of early Universe models.
This project, like so many others that he founded and supported, owes much to the vision and leadership of the late Professor Andrew E. Lange.

\bibliographystyle{aasjournal}
\bibliography{references}

\end{document}